\documentclass[a4paper,11pt,preprint]{elsarticle}

\usepackage{amssymb,hyperref,appendix}
\usepackage{amsmath}

\usepackage[T1]{fontenc} 
\usepackage[lmargin=2.65cm, rmargin=2.65cm, bmargin=3.5cm]{geometry}
\begin{document} 

\begin{frontmatter}
\title{Compton-like scattering of a scalar particle with $N$ photons and one graviton}

%\vspace{1cm}

\author[a]{Naser Ahmadiniaz}
\ead{n.ahmadiniaz@hzdr.de}

\author[b]{Filippo Maria Balli}
\ead{filippo.balli@studenti.unimore.it} 

\author[b,c,d]{Olindo Corradini}
\ead{olindo.corradini@unimore.it}

\author[e]{Jos\'e Manuel D\'avila}
\ead{jmdd@pm.me}

\author[d,f]{Christian Schubert}
\ead{schubert@ifm.umich.mx}

\address[a]{Helmholtz-Zentrum Dresden-Rossendorf, Bautzner Landstra\ss e 400, 01328 Dresden, Germany}
\address[b]{Dipartimento di Scienze Fisiche, Informatiche e Matematiche, Universit\`a degli Studi di Modena e Reggio Emilia, Via Campi 213/A, I-41125 Modena, Italy}
\address[c]{ INFN, Sezione di Bologna,  Via Irnerio 46, I-40126 Bologna, Italy}
\address[d]{Max-Planck-Institut f\"ur Gravitationphysik, Albert-Einstein-Institut, Am M\"uhlenberg 1, 14476 Golm, Germany}
\address[e]{Facultad de Ciencias, Universidad Aut\'onoma del Estado de M\'exico, Instituto Literario 100, C.P. 50000, Toluca, M\'exico}
\address[f]{Instituto de F\'isica y Matem\'aticas
Universidad Michoacana de San Nicol\'as de Hidalgo
Edificio C-3, Apdo. Postal 2-82
C.P. 58040, Morelia, Michoac\'an, Mexico}
% e-mail addresses: one for each author, in the same order as the au

%\cortext[cor1]{Corresponding author.}

\begin{keyword}
Scattering Amplitudes, Gauge symmetry, Space-time symmetries \\
\vspace{0.3cm}
\hspace{9cm}{\it To the memory of Corneliu Sochichiu}
\end{keyword}

\begin{abstract}
Tree-level scattering amplitudes for a scalar particle coupled to an arbitrary number $N$ of photons and a single graviton are computed. We employ the worldline formalism as  the main tool to compute the irreducible part of the amplitude, where all the photons and the graviton are directly attached to the scalar line, then derive a ``tree replacement'' rule to construct the reducible parts of the amplitude which involve irreducible pure $N$-photon two-scalar amplitudes where one photon line emits the graviton. We test our construction by verifying the on-shell gauge and diffeomorphism Ward identities, at arbitrary $N$. 
\end{abstract}

\end{frontmatter}

\tableofcontents

%------------------------------------------------------------------------     
%COLORS
\def\green{\color{green}}
\def\blue{\color{blue}}
\def\red{\color{red}}
\def\black{\color{black}}
% MATH SYMBOLS
%

\def\veps{\varepsilon}
\newcommand{\Zz}{\mathcal{Z}}
\newcommand{\Zzp}{\mathcal{Z}^{\prime}}
\newcommand{\detZ}{\textrm{det}^{-\frac{1}{2}}\left[\frac{\sin\Zz}{\mathcal{\Zz}}\right]}
\newcommand{\detZp}{\textrm{det}^{-\frac{1}{2}}\left[\frac{\sin \Zzp}{\Zzp}\right]}
\newcommand{\detZs}{\textrm{det}^{-\frac{1}{2}}\left[\frac{\tan\Zz}{\mathcal{\Zz}}\right]}
\newcommand{\detZps}{\textrm{det}^{-\frac{1}{2}}\left[\frac{\tan\Zzp}{\Zzp}\right]}
\newcommand{\pdetZ}{\textrm{det}^{-\frac{1}{2}}\left[\cos \Zz \right]}
\newcommand{\pdetZp}{\textrm{det}^{-\frac{1}{2}}\left[\cos(\Zzp)\right]}

\newcommand{\tZz}{\frac{\tan\Zz}{\Zz}}
\newcommand{\tZzp}{\frac{\tan\Zzp}{\Zzp}}
\newcommand{\link}{\Big\vert_k}

\newcommand{\xm}{x_{-}}
\newcommand{\xp}{x_{+}}
\newcommand{\yp}{y_{+}}
\newcommand{\ym}{y_{-}}

\newcommand{\delC}{\underset{\smile}{\Delta}}
\newcommand{\ddelC}{{^{\bullet}\!\delC}}
\newcommand{\delCd}{{\delC\!^{\bullet}}}
\newcommand{\ddelCd}{{^{\bullet}\!\delC\!^{\bullet}}}
\newcommand{\odelC}{{^{\circ}\!\delC}}
\newcommand{\delCo}{{\delC\!^{\circ}}}
\newcommand{\odelCo}{{^{\circ}\!\delC\!^{\circ}}}
\newcommand{\odelCd}{{^{\circ}\!\delC\!^{\bullet}}}
\newcommand{\ddelCo}{{^{\bullet}\!\delC\!^{\circ}}}

\newcommand{\gb}{{\mathcal{G}_{B}}}
\newcommand{\gbd}{{\dot{\mathcal{G}}_{B}}}
\newcommand{\gbdm}{\dot{\mathcal{G}}_{B \mu\nu}}
\newcommand{\gf}{{\mathcal{G}_{F}}}
\newcommand{\gfd}{{\dot{\mathcal{G}}_{F}}}
\newcommand{\gfdm}{\dot{\mathcal{G}}_{F \mu\nu}}

\def\cZ{{\cal Z}}

\def\cosech{\rm cosech}
\def\sech{\rm sech}
\def\coth{\rm coth}
\def\tanh{\rm tanh}
\def\tan{\rm tan}
%fractions
\def\half{{1\over 2}}
\def\third{{1\over3}}
\def\fourth{{1\over4}}
\def\fifth{{1\over5}}
\def\sixth{{1\over6}}
\def\seventh{{1\over7}}
\def\eigth{{1\over8}}
\def\ninth{{1\over9}}
\def\tenth{{1\over10}}
\def\conj{{{\rm c.c.}}}
\def\bN{\mathop{\bf N}}
\def\R{{\rm I\!R}}
\def\Eins{{\mathchoice {\rm 1\mskip-4mu l} {\rm 1\mskip-4mu l}
{\rm 1\mskip-4.5mu l} {\rm 1\mskip-5mu l}}}
\def\Z{{\mathchoice {\hbox{$\sf\textstyle Z\kern-0.4em Z$}}
{\hbox{$\sf\textstyle Z\kern-0.4em Z$}}
{\hbox{$\sf\scriptstyle Z\kern-0.3em Z$}}
{\hbox{$\sf\scriptscriptstyle Z\kern-0.2em Z$}}}}
\def\abs#1{\left| #1\right|}
\def\com#1#2{
        \left[#1, #2\right]}
%\def\square{\kern1pt\vbox{\hrule height 1.2pt\hbox{\vrule width 1.2pt
%   \hskip 3pt\vbox{\vskip 6pt}\hskip 3pt\vrule width 0.6pt}
%   \hrule height 0.6pt}\kern1pt}
 %     \def\boxop{{\raise-.25ex\hbox{\square}}}
% \contract is a differential geometry contraction sign _|
\def\contract{\makebox[1.2em][c]{
        \mbox{\rule{.6em}{.01truein}\rule{.01truein}{.6em}}}}
\def\ltap{\ \raisebox{-.4ex}{\rlap{$\sim$}} \raisebox{.4ex}{$<$}\ }
\def\gtap{\ \raisebox{-.4ex}{\rlap{$\sim$}} \raisebox{.4ex}{$>$}\ }
\def\mn{{\mu\nu}}
\def\rs{{\rho\sigma}}
\newcommand{\Det}{{\rm Det}}
\def\Tr{{\rm Tr}\,}
\def\tr{{\rm tr}\,}
\def\sumij{\sum_{i<j}}
\def\e{\,{\rm e}}
%boldface vectors
\def\br{{\bf r}}
\def\bp{{\bf p}}
\def\bq{{\bf q}}
\def\bx{{\bf x}}
\def\by{{\bf y}}
\def\brhat{{\bf \hat r}}
\def\bv{{\bf v}}
\def\ba{{\bf a}}
\def\bE{{\bf E}}
\def\bB{{\bf B}}
\def\bA{{\bf A}}
\def\b0{{\bf 0}}
%derivatives
\def\pa{\partial}
\def\dA{\partial^2}
\def\ddx{{d\over dx}}
\def\ddt{{d\over dt}}
\def\der#1#2{{d #1\over d#2}}
\def\lie{\hbox{\it \$}} % fancy L for the Lie derivative
\def\partder#1#2{\frac{\partial #1}{\partial #2}}
\def\secder#1#2#3{{\partial^2 #1\over\partial #2 \partial #3}}
%
%equations
%\newcommand{\be}{\blue\begin{equation}}
%\newcommand{\ee}{\end{equation}\black\noindent}
%\newcommand{\bear}{{\blue\begin{eqnarray}}}
%\newcommand{\ear}{{\end{eqnarray}\black\noindent}}
%\newcommand{\benn}{\begin{enumerate}}
%\newcommand{\enn}{\end{enumerate}}
%\newcommand{\veject}{\vfill\eject}
%\newcommand{\ven}{\vfill\eject\noindent}
\def\be{\begin{equation}}
\def\ee{\end{equation}\noindent}
\def\bear{\begin{eqnarray}}
\def\ear{\end{eqnarray}\noindent}
\def\bec{\blue\begin{equation}}
\def\eec{\end{equation}\black\noindent}
\def\bearc{\blue\begin{eqnarray}}
\def\earc{\end{eqnarray}\black\noindent}
\def\benn{\begin{enumerate}}
\def\enn{\end{enumerate}}
\def\veject{\vfill\eject}
\def\ven{\vfill\eject\noindent}
%
%reference to equations
\def\eq#1{{eq. (\ref{#1})}}
\def\eqs#1#2{{eqs. (\ref{#1}) -- (\ref{#2})}}
%
%algebra
\def\inv#1{\frac{1}{#1}}
\def\sumninf{\sum_{n=0}^{\infty}}
%
%integrals
\def\totint{\int_{-\infty}^{\infty}}
\def\posint{\int_0^{\infty}}
\def\negint{\int_{-\infty}^0}
\def\pint{{\dps\int}{dp_i\over {(2\pi)}^d}}
\def\intdp3{\int\frac{d^3p}{(2\pi)^3}}
\def\intdp4{\int\frac{d^4p}{(2\pi)^4}}
%propagators
\def\scalprop#1{\frac{-i}{#1^2+m^2-i\epsilon}}
%
% PHYS SYMBOLS
\newcommand{\GeV}{\mbox{GeV}}
\def\FFdual{F\cdot\tilde F}
\def\bra#1{\langle #1 |}
\def\ket#1{| #1 \rangle}
\def\braket#1#2{\langle {#1} \mid {#2} \rangle}
\def\vev#1{\langle #1 \rangle}
\def\matel#1#2#3{\langle #1\mid #2\mid #3 \rangle}
\def\rightvac{\mid0\rangle}
\def\leftvac{\langle0\mid}
\def\ihbar{{i\over\hbar}}
\def\lagr{{\cal L}}
% spinor stuff
\def\sigmabar{{\bar\sigma}}
% dirac matrix stuff
\def\ge{\hbox{$\gamma_1$}}
\def\gz{\hbox{$\gamma_2$}}
\def\gd{\hbox{$\gamma_3$}}
\def\go{\hbox{$\gamma_1$}}
\def\gt{\hbox{$\gamma_2$}}
\def\gth{\hbox{$\gamma_3$}} 
\def\gf{\hbox{$\gamma_5\;$}}
\def\slash#1{#1\!\!\!\raise.15ex\hbox {/}}
\newcommand{\slD}{\,\raise.15ex\hbox{$/$}\kern-.27em\hbox{$\!\!\!D$}}
\newcommand{\slpartial}{\raise.15ex\hbox{$/$}\kern-.57em\hbox{$\partial$}}
\newcommand{\PP}{\cal P}
\newcommand{\G}{{\cal G}}
\newcommand{\nc}{\newcommand}
\nc{\spa}[3]{\left\langle#1\,#3\right\rangle}
\nc{\spb}[3]{\left[#1\,#3\right]}
\nc{\ksl}{\not{\hbox{\kern-2.3pt $k$}}}
\nc{\hf}{\textstyle{1\over2}}
\nc{\pol}{\varepsilon}
\nc{\tq}{{\tilde q}}
\nc{\esl}{\not{\hbox{\kern-2.3pt $\pol$}}}
\newcommand{\cL}{\cal L}
\newcommand{\D}{\cal D}
\newcommand{\Dhalf}{{D\over 2}}
\def\eps{\epsilon}
\def\epshalf{{\epsilon\over 2}}
\def\lag{( -\partial^2 + V)}
%worldline
\def\freeexp{{\rm e}^{-\int_0^Td\tau {1\over 4}\dot x^2}}
\def\kinb{{1\over 4}\dot x^2}
\def\kinf{{1\over 2}\psi\dot\psi}
\def\expk{{\rm exp}\biggl[\,\sum_{i<j=1}^4 G_{Bij}k_i\cdot k_j\biggr]}
\def\expp{{\rm exp}\biggl[\,\sum_{i<j=1}^4 G_{Bij}p_i\cdot p_j\biggr]}
\def\expshort{{\e}^{\half G_{Bij}k_i\cdot k_j}}
\def\expabb{{\e}^{(\cdot )}}
\def\epseps#1#2{\varepsilon_{#1}\cdot \varepsilon_{#2}}
\def\epsk#1#2{\varepsilon_{#1}\cdot k_{#2}}
\def\kk#1#2{k_{#1}\cdot k_{#2}}
\def\G#1#2{G_{B#1#2}}
\def\Gp#1#2{{\dot G_{B#1#2}}}
\def\GF#1#2{G_{F#1#2}}
\def\Dab{{(x_a-x_b)}}
\def\Dsq{{({(x_a-x_b)}^2)}}
\def\PITD{{(4\pi T)}^{-{D\over 2}}}
\def\4piTD{{(4\pi T)}^{-{D\over 2}}}
\def\4piT4{{(4\pi T)}^{-2}}
\def\TintmD{{\dps\int_{0}^{\infty}}{dT\over T}\,e^{-m^2T}
    {(4\pi T)}^{-{D\over 2}}}
\def\Tintm4{{\dps\int_{0}^{\infty}}{dT\over T}\,e^{-m^2T}
    {(4\pi T)}^{-2}}
\def\Tintm{{\dps\int_{0}^{\infty}}{dT\over T}\,e^{-m^2T}}
\def\Tint{{\dps\int_{0}^{\infty}}{dT\over T}}
\def\np{n_{+}}
\def\nm{n_{-}}
\def\Np{N_{+}}
\def\Nm{N_{-}}
\newcommand{\slG}{{{\dot G}\!\!\!\! \raise.15ex\hbox {/}}}
\newcommand{\Gd}{{\dot G}}
\newcommand{\Gund}{{\underline{\dot G}}}
\newcommand{\Gdd}{{\ddot G}}
\def\GBd12{{\dot G}_{B12}}
\def\Dx{\dps\int{\cal D}x}
\def\Dy{\dps\int{\cal D}y}
\def\Dpsi{\dps\int{\cal D}\psi}
\def\dint#1{\int\!\!\!\!\!\int\limits_{\!\!#1}}
\def\ddtau{{d\over d\tau}}
\def\ie{\hbox{$\textstyle{\int_1}$}}
\def\iz{\hbox{$\textstyle{\int_2}$}}
\def\id{\hbox{$\textstyle{\int_3}$}}
\def\ldop{\hbox{$\lbrace\mskip -4.5mu\mid$}}
\def\rdop{\hbox{$\mid\mskip -4.3mu\rbrace$}}
%
%VARIOUS
\newcommand{\1}{{\'\i}}
\newcommand{\no}{\noindent}
\def\non{\nonumber}
\def\dps{\displaystyle}
\def\sy{\scriptscriptstyle}
\def\sy{\scriptscriptstyle}

\section{Introduction}
\label{sec:intro}

The systematic computation of various classes of on-shell scattering amplitudes has become a very active field of research in the past few decades, and several very efficient methods have been put forward, which involve spinor helicity formalism,  on-shell recursion relations, Ward identities, KLT relations, just to name a few---see~\cite{Elvang:2015rqa} for a review. The main paradigm consists in avoiding the use of lagrangian field theories with all their plethoric structures and rely instead on more basic features (symmetries, kinematics,...) that allow to more efficiently compute the amplitudes. One feature which is common to most of these very effective methods---particularly those where the spinor helicity tricks are used as the main tool---is the masslessness of the propagating particles. As a complementary method in the present manuscript we consider a worldline approach, which instead ideally works with massive propagating particles. 

Historically, the first pioneering work on the worldline approach to Quantum Field Theory is due to Feynman, who proposed a particle path integral representation for the dressed propagator of a scalar field coupled to electromagnetism~\cite{Feynman:1951gn}. However, this formulation
was not taken seriously as an alternative to the standard Feynman diagram method 
for the actual computation of effective actions and scattering amplitudes until the early nineties,
when Bern and Kosower~\cite{Bern:1990cu, Bern:1991aq} 
derived novel rules for the construction of one-loop $N$ - gluon amplitudes from 
first-quantized open string theory, and similar rules were shortly later derived from
the closed string for one-loop $N$ - graviton amplitudes \cite{bedush}. 

For the gluonic case, these rules were then rederived 
from point particle path integrals by Strassler~\cite{Strassler:1992zr}, which established
this ``worldline formalism'' as a serious alternative to Feynman diagrams, and 
triggered a host of generalizations to other types of amplitudes and effective actions 
---see Ref.~\cite{Schubert:2001he} for an earlier account on the development of the method. So far, the majority of developments and applications of the worldline approach have been at the loop level:
multiloop calculations~\cite{8,Schmidt:1994aq}, worldline methods with strong external fields~\cite{17,Reuter:1996zm, Gies:2005sb, Dunne:2005sx},  the worldline formalism in curved spacetime~\cite{Bastianelli:2002fv}, one-loop quantum gravity~\cite{Bastianelli:2013tsa, Bastianelli:2019xhi} and photon-graviton mixing in an electromagnetic field~\cite{Bastianelli:2004zp}, the worldline Monte-Carlo approach to the Casimir effect~\cite{Gies:2003cv}, higher-spin field theory~\cite{Bastianelli:2007pv, Bastianelli:2012bn}, and applications to QFT on manifolds with boundary~\cite{Bastianelli:2006hq, Corradini:2019nbb}, noncommutative QFT's~\cite{Bonezzi:2012vr} and form-factor decompositions of off-shell gluon amplitudes \cite{Ahmadiniaz:2012xp,Ahmadiniaz:2016qwn} and many more. 

On other hand, the worldline approach to dressed propagators and to the associated scattering amplitudes is a much less developed subject of research, though the Bern-Kosower rules for a scalar particle line coupled to electromagnetism in vacuum were found soon after their one-loop counterparts~\cite{Daikouji:1995dz, Ahmadiniaz:2015kfq}. However, more recently,   master formulas for a scalar particle in a constant background field were derived~\cite{Ahmad:2016vvw}, and the coupling to non-abelian fields in vacuum was also studied~\cite{Ahmadiniaz:2015xoa}. Generalizations to propagators of fields with spin are even more rare. The straightforward procedure would be to consider locally supersymmetric spinning particle models on the worldline~\cite{Gershun:1979fb}---recently Einstein gravity was studied through the BRST quantization of an ${\cal N}=4$ spinning particle model~\cite{Bonezzi:2018box}. However, there are technical difficulties to be overcome in the path integral quantization of such models on the open line, since the gravitino present in the locally supersymmetric model cannot be completely gauged away, and the coherent state boundary conditions for the fermionic coordinates, responsible for providing the spinorial degrees of freedom to the particle, do not appear to be very convenient. A suitable alternative approach is to employ the `Symb' map developed in~\cite{brezinmarinov-77,Fradkin:1991ci} which reproduces the spin-factor potential in terms of fermionic coordinates with antiperiodic boundary conditions, and the resulting particle models are now globally supersymmetric. This approach allowed to compute some previously neglected one-particle reducible contributions to the fermion propagator in a constant field~\cite{Ahmadiniaz:2017rrk}. Moreover, a derivation of a master formula for the tree level $N$-photon fermion propagator is under completion~\cite{fppaper1}. 

In the present manuscript we instead take a path towards the derivation of tree level amplitudes with gravitons, using as a main tool the worldline approach in curved space. In fact, at the level of one photon---i.e. for the gravitational photoproduction process---the amplitude displays a very interesting factorization property ~\cite{geohal-81, chshso-95, Holstein:2006ry, Bastianelli:2012bz, Bjerrum-Bohr:2014lea, Ahmadiniaz:2016vai}, which is briefly reviewed below in a dedicated subsection. However, this nice factorization property appears not to work beyond the $N=1$ case since there are too few
conservation laws---see~\cite{geohal-81,chshso-95}  for a detailed discussion.
%
%{\color{red} However, this nice factorization property appears not to work beyond the {\color{blue} $N=1$} \st{four-point case since there are only three} {\color{blue} due to the lack of enough} conservation laws---see \cite{geohal-81, chshso-95} for detailed discussion.} 
%
Here, we consider a scalar particle line perturbatively coupled to electromagnetism and to gravity, and provide a master formula which involves the inclusion of $N$ photons and one graviton into the scalar line, i.e. we add a graviton to Daikouji et al's formula~\cite{Daikouji:1995dz} which was also rederived in \cite{Ahmadiniaz:2015kfq}. The inclusion even of a single graviton is by no means trivial for various reasons. Firstly, it boils down to the application of the worldline formalism in curved space which, although being well understood by now, is certainly trickier than its flat space counterpart. In fact, the coupling to gravity, in the perturbative approach requires the use of regularization schemes and careful treatment of the non-trivial path integral measure~\cite{Bastianelli:2006rx}. Moreover, the graviton can either couple directly  to the scalar line, but it may also be emitted from a photon line, since gravity couples to the photon stress tensor. This second  contribution involves diagrams that are one-particle reducible in the photon lines, and akin to what occurs in the presence of a non-abelian gauge field where, say, a gluon emitted from the scalar line can split in two or three gluons. Similar issues were indeed already discussed, for instance in the worldgraph approach to Yang-Mills amplitudes~\cite{Dai:2008bh} and in  worldline calculations~\cite{FMB-thesis}. 

A particularly elegant feature of the original Bern-Kosower and Bern-Dunbar-Shimada rules
for gluon and graviton amplitudes is that they provide a simple rule for constructing the reducible
contributions from the irreducible ones at the integrand level, instead of the usual ``sewing trees onto
loops'' procedure. 
Here, we provide a similar novel replacement rule, which allows us to obtain the reducible part of the amplitudes with the graviton in terms of the scalar lines with only photons attached, thus in terms of amplitudes for which a convenient generating master formula exists. For this purpose, it will be essential that in the worldline approach to scattering amplitudes there is a priori no need to impose on-shell conditions on the external lines.

In the following we first rederive the $N$-photon scalar propagator and the associated master formula, since it is one of our main tools. Then we consider the insertion of a graviton and single out the irreducible part of the amplitudes---by using a helpful parametrization of the graviton polarization, and the reducible part through the aforementioned replacement rule. This allows us to give a compact formula for the full tree level amplitude with $N$ photons, one graviton and two scalars. 
%{\color{red} fully off-shell, i.e. scalar, photons and gravitons are kept off-shell}.
%We use it to explicitly compute the $N=0,1,2$ cases. 
We thus test our master formula by checking the on-shell transversality on the photon lines and graviton lines. In the graviton case, this requires a conspiration between the reducible and irreducible
contributions that becomes rather transparent in our approach. Some computational details, concerning amplitudes with $N\leq 2$, are
relegated to the appendix.

%%%%%%%%%%%%%%%%%%%%%%%%%%%%%%%%%%%%%%%%%%%%%%%%%%%%%%%%%%
\section{$N$-photon scalar propagator from the worldline formalism}
\label{sec:photon-ampl}

The photon-dressed propagator in scalar QED can be efficiently obtained using the line path integral of a scalar particle in the presence of an external electromagnetic field,
\begin{align}
\Big\langle \phi(x') \bar \phi(x)\Big\rangle_A = \int_0^\infty dT e^{-m^2T}\int_{x(0)=x}^{x(T)=x'}Dx~e^{-\int_0^T d\tau\big(\frac1{4}\dot x^2+ie\dot x\cdot A(x)\big)}~.
\label{eq:propx}
\end{align}
The $N$-photon scalar propagator, i.e. the scalar propagator with the insertion of $N$ photons can be obtained
with the straightforward recipe that we briefly review here. Firstly, write the external field as a sum of $N$ photons
\begin{align}
	A_\mu(x(\tau))= \sum_{l=1}^N\varepsilon_{l,\mu} e^{ik_l\cdot x}\,,
\end{align}
then extract from~\eqref{eq:propx} the multi-linear part, in the various polarizations $\varepsilon_l$, and Fourier transform in the two external scalar lines. This leads to
\begin{align}
	D^{(N)}(p,p';\varepsilon_1,k_1,\dots, \varepsilon_N,k_N)&=(-ie)^N \int_0^\infty dT e^{-m^2T}\int d^4x\int d^4x' e^{i(p\cdot x+p'\cdot x')}\nonumber\\&
	\times \int_{x(0)=x}^{x(T)=x'}Dx~e^{-\int_0^Td\tau\frac1{4}\dot x^2}\prod_{l=1}^N\int_0^T d\tau_l \, \varepsilon_l\cdot \dot x(\tau_l) e^{ik_l\cdot x(\tau_l)}~.
\end{align}
It is thus convenient to split the particle path in terms of a background $\bar x^\mu(\tau)=x^\mu +(x'^\mu-x^\mu)\tfrac{\tau}{T}$ and fluctuations $q^\mu(\tau)$ with vanishing boundary conditions. One thus gets
\begin{align}
	&D^{(N)}(p,p'; \varepsilon_1,k_1,\dots, \varepsilon_N,k_N)=(-ie)^N \int_0^\infty dT e^{-m^2T}\int d^4x\int d^4x' e^{i(p\cdot x+p'\cdot x')-\frac1{4T}(x-x')^2} \nonumber\\&
	\times e^{\sum_l \big(ik_l\cdot x+\tfrac{\varepsilon_l}{T}\cdot(x'-x)\big)}\int_{q(0)=0}^{q(T)=0}Dq~e^{-\int_0^T d\tau\frac1{4}\dot q^2} \prod_{l=1}^N\int_0^Td\tau_l \, e^{ik_l\cdot \big((x'-x)\tfrac{\tau_l}{T}+q(\tau_l)\big)+\varepsilon_l\cdot \dot q(\tau_l)}\Biggr|_{\rm m.l.}~,
\end{align}
where 'm.l.' indicates that we are only meant to pick out the multilinear part in all the polarizations. The latter path integral thus provides the correlation function of the product of $N$ photon vertex operators
\begin{align}
	V_A[\varepsilon,k]=e^{ik\cdot x+\tfrac{\varepsilon}{T}\cdot(x'-x)}\int_0^Td\tau \,e^{ik\cdot \big((x-x')\tfrac{\tau}{T}+q(\tau)\big)+\varepsilon\cdot \dot q(\tau)}\Biggr|_{\rm lin}\,,
\end{align}
with respect to the Gaussian measure $\int Dq \,e^{-\frac1{4}\int \dot q^2}$, which has normalization $\frac1{(4\pi T)^{D/2}}$ and yields the Green's functions
\begin{align}
	&\big\langle q^\alpha(\tau) q^{\alpha'}(\tau')\big\rangle = -2\delta^{\alpha\alpha'}\Delta(\tau,\tau')\,,\\
	&\label{eq:delta} \Delta(\tau,\tau')=\frac{\tau \tau'}{T}+\frac12 |\tau-\tau'|-\frac12(\tau+\tau')~.
\end{align}
Thus, after some straightforward algebra one finds the Bern-Kosower-like master formula originally obtained by Daikouji et al~\cite{Daikouji:1995dz} and later in the worldline formalism in \cite{Ahmadiniaz:2015kfq}, i.e.
\begin{align}
&\widetilde D^{(N)}(p,p'; \varepsilon_1,k_1,\dots, \varepsilon_N,k_N)=(-ie)^N\int_0^\infty dT e^{-T(m^2+p'^2)}\prod_{l=1}^N\int_0^T d\tau_l\nonumber\\
&\times\exp\Big\{(p'-p)\cdot \sum_{l=1}^N(-k_l\tau_l +i\varepsilon_l	) +\sum_{l,l'=1}^N\big( k_l\cdot k_{l'}\Delta_{l-l'}-2i\varepsilon_l\cdot k_{l'}\dot \Delta_{l-l'} +\varepsilon_l\cdot \varepsilon_{l'} \ddot \Delta_{l-l'}\big) \Big\}\Biggr|_{\rm m.l.}~,
\label{eq:tildeA}
\end{align}
where
\begin{align}
\Delta_{l-l'} := \frac12 |\tau_l-\tau_{l'}|\,,
\end{align}
is the translation-invariant part of~\eqref{eq:delta}. Above we have also stripped off the overall momentum-conservation delta function. The Feynman amplitude for the tree-level scattering  of two scalars and $N$ photons can thus be obtained from~\eqref{eq:tildeA} by truncating the external scalar lines, i.e. multiplying by $(p^2+m^2)(p'^2+m^2)$,
\begin{align}
{\cal D}^{(N)} (p,p'; \varepsilon_1,k_1,\dots, \varepsilon_N,k_N)= (p^2+m^2)(p'^2+m^2)\widetilde D^{(N)}(p,p'; \varepsilon_1,k_1,\dots, \varepsilon_N,k_N)~.
\label{eq:MA}
\end{align}
Note that, as already mentioned in the Introduction, this expression holds off the mass-shell of the external particles. However, going on-shell leads to transversality in all the photon lines, upon the replacement $\varepsilon_l(k_l)\ \to\ k_l$, as will be reviewed below. %\footnote{\color{red} To avoid any source of misunderstanding and confusion note that the notation we use for fully off-shell amplitude $\cal{M}^{(N)}$ might coincides with the one of on-shell Feynman amplitude in text books.}. 

%%%%%%%%%%%%%%%%%%%%%%%%%%%%%%%%%
\section{Insertion of a graviton}
\label{sec:gravi-ampl}

The computation of scattering amplitudes of the scalar particle with photons and gravitons, can be performed by considering the worldline representation in curved space~\cite{Bastianelli:2000nm}. For the propagator of a scalar particle minimally coupled to gravity, we have~\footnote{Here `minimally coupled' refers to the minimal coupling in the worldline action, i.e. $\bar\xi=\xi-\tfrac14=0$, which renders the graviton vertex operator linear in $\epsilon_{\mu\nu}$, as opposed to the minimal coupling in the spacetime action, which corresponds to $\xi=0$. }
\begin{align}
	\Big\langle \phi(x') \bar \phi(x)\Big\rangle_{A,g}&= \int_0^\infty dT e^{-m^2T}\int_{x(0)=x}^{x(T)=x'}DxDaDbDc \nonumber \\ 
	&\times e^{-\int_0^Td\tau\big(\frac1{4}g_{\mu\nu}(x)(\dot x^\mu \dot x^\nu +a^\mu a^\nu+b^\mu c^\nu)+ie\dot x\cdot A(x)\big)}\,,
	\label{eq:dressed-prop}
\end{align}
where the fields $a^\mu$, and $b^\mu$ and $c^\mu$ are commuting, respectively anti-commuting, auxiliary fields (Lee-Yang ghosts) which were found to suitably represent the Einstein-invariant path integral measure, and have vanishing boundary conditions. By expanding the metric about the flat background
\begin{align}
g_{\mu\nu}(x) =\delta_{\mu\nu} +\kappa \epsilon_{\mu\nu}e^{ik_0\cdot x}\,,
\end{align}
and using the same split described above for the particle paths---we can read off the graviton vertex operator 
\begin{align}
V_g[\epsilon,k_0]=e^{ik_0\cdot x+\frac1{T^2}(x'-x)\cdot\epsilon\cdot(x'-x)}\int_0^Td\tau \,e^{ik_0\cdot \big((x'-x)\frac{\tau}{T}+q\big)+\epsilon_{\mu\nu}\big( \frac2T (x'-x)^\mu \dot q^\nu+\dot q^\mu \dot q^\nu +a^\mu a^\nu +b^\mu c^\nu\big)} \Biggr|_{\rm lin}\,,
\label{eq:grav-vert-op}
\end{align} 
along with auxiliary fields propagators
\begin{align}
	&\big\langle a^\mu(\tau)a^\nu(\tau')\big\rangle =2\delta^{\mu\nu}\delta(\tau,\tau')\,,\\
	&\big\langle b^\mu(\tau)c^\nu(\tau')\big\rangle =-4\delta^{\mu\nu}\delta(\tau,\tau')~.
\end{align}
Hence, the irreducible part of the tree-level scalar propagator with the insertion of $N$ photons and one graviton reads
\begin{align}
	& D^{(N,1)}(p,p'; \varepsilon_1,k_1,\dots, \varepsilon_N,k_N;\epsilon,k_0) =(-ie)^N \left(-\frac{\kappa}{4}\right)\int_0^\infty dT e^{-m^2T}\nonumber\\&\times\int d^4x\int d^4x'  %\nonumber\\ & 
	e^{i(p\cdot x+p'\cdot x')-\frac1{4T}(x-x')^2} \frac1{(4\pi T)^\frac D2} \, \Bigg\langle\prod_{l=1}^N V_A[\varepsilon_l,k_l]\, V_g[\epsilon,k_0]\Bigg\rangle\,,
\end{align}
where only the part linear in all the polarizations ($\varepsilon$'s and $\epsilon$) has to be retained. In the next sections we provide a specific recipe to handle this task and obtain a useful master formula for the full Feynman amplitude.

%%%%%%%%%%%%%%%%%%%%%%%%%%%%%%%%%%
\subsection{Irreducible part of the amplitude}
\label{sec:irred-part}
\begin{figure}[htbp]
\begin{center}
 \includegraphics[width=0.7\textwidth]{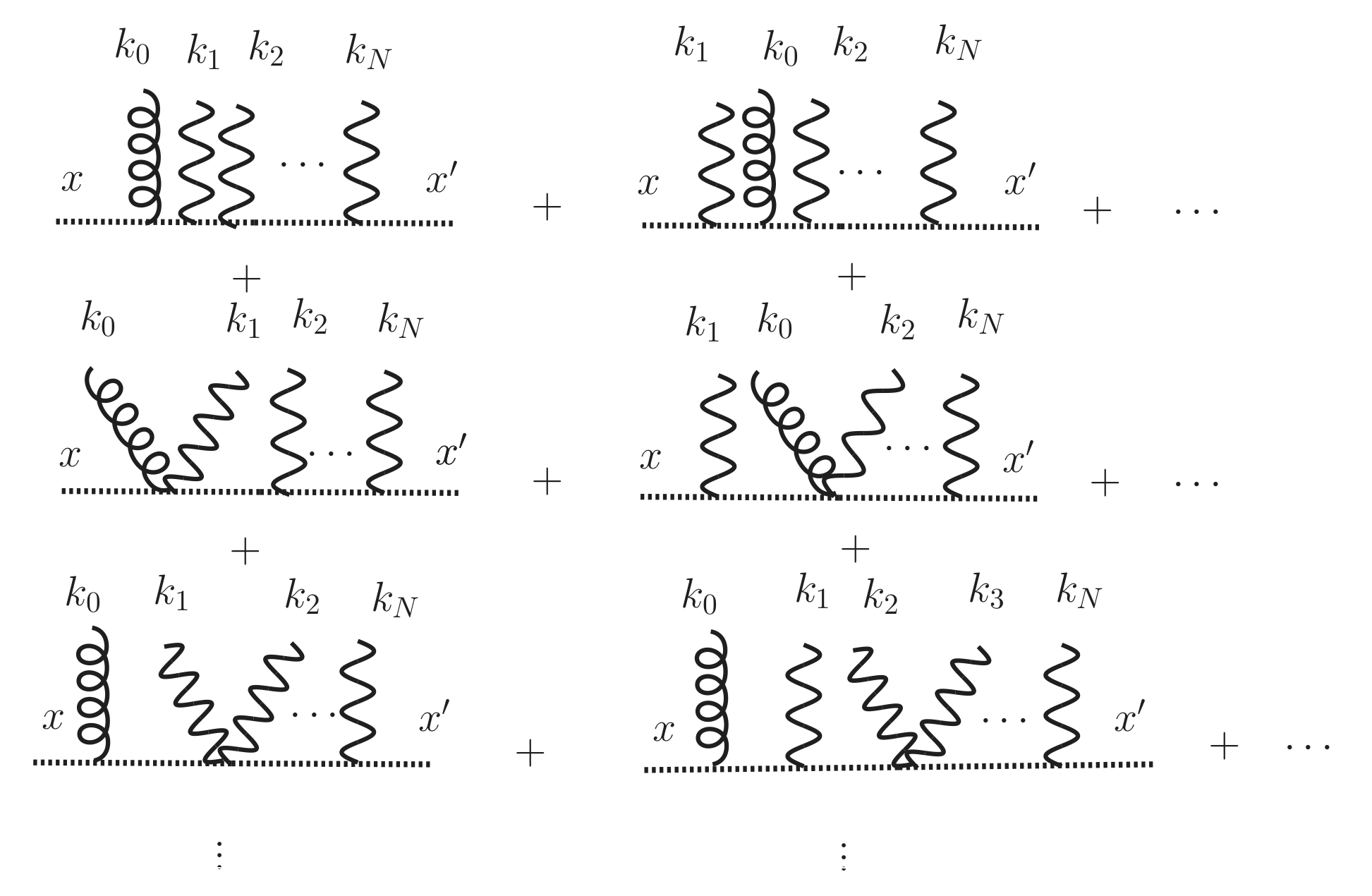}
\caption{The Feynman diagram representation (in configuration space) for irreducible contributions to $N$-photon one-graviton amplitude. The diagrams in the second and third lines involve quartic vertices that in the worldline approach come from delta functions, e.g. the first one is given by $\delta(\tau_0-\tau_1)$ etc.}
\label{1grNph-irr}
\end{center}
\end{figure}

In order to explicitly compute the irreducible part of the $N$-photon one-graviton amplitude (see Fig. \ref{1grNph-irr}) we find it convenient to parametrize the graviton polarization as
\begin{align}
&\epsilon_{\mu\nu} := \lambda_{\mu} \rho_{\nu}\,,
\label{eq:e}\\
& \varepsilon_{0\mu} := \lambda_{\mu} + \rho_{\mu}~,
\label{eq:lr}
\end{align} 
where, in Eq.~\eqref{eq:e}, symmetrization between indices is implied. 
Such parametrization has to be understood as a simple book-keeping device to combine photon and graviton insertions together; at the end the graviton polarization is reconstructed from the term simultaneously linear in $\lambda$ and $\rho$.
In fact, with a single graviton insertion, the ghost contribution cancels against the singular part of the $\langle \dot q^\mu(\tau_0) \dot q^\nu(\tau_0) \rangle$ propagator that appears in the graviton vertex operator. We can thus neglect the ghost contributions, provided we take $\langle \dot q^\mu(\tau_0) \dot q^\nu(\tau_0) \rangle \cong -\frac{2}{T} \delta^{\mu\nu}$ in the graviton sector. The graviton vertex operator can thus be written as
  \begin{align}
V_g[\epsilon,k_0]=e^{ik_0\cdot x+\frac{\varepsilon_0}{T}\cdot(x'-x)}\int_0^Td\tau_0 \,e^{ik_0\cdot \big((x'-x)\frac{\tau_0}{T}+q(\tau_0)\big)+\varepsilon_0\cdot \dot q(\tau_0) }\Big|_{\rm lin.\, \lambda,\, \rho}\,,
\end{align}  
which has the same form as the photon counterpart, with the only subtlety that the linear part in $\lambda$ and $\rho$ comes from the quadratic part in $\varepsilon_0$. We thus get the ``$N$-photon one-graviton scalar propagator''
\begin{align}
&\widetilde D^{(N,1)}(p,p'; \varepsilon_1,k_1,\dots, \varepsilon_N,k_N;\epsilon,k_0)=(-ie)^N\left(-\frac{\kappa}{4}\right)\int_0^\infty dT e^{-T(m^2+p'^2)}\prod_{l=0}^N\int_0^T d\tau_l\nonumber\\
&\times\exp\Big\{(p'-p)\cdot \sum_{l=0}^N(-k_l\tau_l +i\varepsilon_l	) +\sum_{l<l'=0}^N\Big( k_l\cdot k_{l'}|\tau_l-\tau_{l'}|+i(\varepsilon_{l'}\cdot k_{l}-\varepsilon_{l}\cdot k_{l'}){\rm sgn}(\tau_l-\tau_{l'})\nonumber\\&\hskip8cm +2\varepsilon_l\cdot \varepsilon_{l'}\, \delta (\tau_l-\tau_{l'}) \Big) \Big\}\Bigr|_{\rm m.l.}~,
\label{eq:tildeAg}
\end{align}
where `m.l.' stands for `multilinear' i.e. linear in all $\varepsilon_l$, $l=1,...,N$ and linear in $\lambda$ and $\rho$, and with $\ddot \Delta_{0-0'} =0$. On the mass shell of the scalar particle, upon truncation of the external scalar lines, the latter provides a contribution to the tree-level amplitude with $N$ photons, one graviton and two scalars that we will refer to as `irreducible' 
\begin{align}
&{\cal D}^{(N,1)}_{irred} (p,p'; \varepsilon_1,k_1,\dots, \varepsilon_N,k_N;\epsilon,k_0) \nonumber \\&=
(p^2+m^2)(p'^2+m^2) \widetilde D^{(N,1)}(p,p'; \varepsilon_1,k_1,\dots, \varepsilon_N,k_N;\epsilon,k_0)~,
\end{align}
meaning that it cannot be divided into two subdiagrams by cutting a photon line or the graviton line.
 
 Let us single out some special cases of the previous formula which will be helpful later. Let us begin considering the case $N=0$, i.e. the `graviton-scalar' vertex,
\begin{align}
&\widetilde D^{(0,1)}(p,p';\epsilon,k_0)=\left(-\frac{\kappa}{4}\right)\int_0^\infty dT~e^{-T(m^2+p'^2)}\int_0^T d\tau_0%\nonumber\\&\times
~e^{(p'-p)\cdot(-k_0{\tau_0} +i\varepsilon_0	)}\Big|_{\rm m.l.}~, 
\end{align}
which, using momentum conservation, can be reduced to
\begin{align}
\widetilde D^{(0,1)}(p,p';\epsilon,k_0)=\frac{\kappa}{4} (p'-p)^\mu \epsilon_{\mu\nu}  (p'-p)^\nu \frac{1}{(p'^2+m^2)(p^2+m^2)}\,,
\end{align} 
and, upon truncation, leads to the amplitude (vertex)
\begin{align}
{\cal D}^{(0,1)} (p,p';\epsilon,k_0)= \frac{\kappa}{4} (p'-p)^\mu \epsilon_{\mu\nu}  (p'-p)^\nu~.
\label{eq:g}
\end{align}
For $N=1$, the irreducible part of the gravitational photoproduction amplitude can be easily obtained from
\begin{align}
&\widetilde D^{(1,1)}(p,p'; \varepsilon_1,k_1;\epsilon,k_0)=(-ie)\left(-\frac{\kappa}{4}\right)\int_0^\infty dT e^{-T(m^2+p'^2)}\int_0^T d\tau_0 \int_0^T d\tau_1\nonumber\\
&\times e^{(p'-p)\cdot (-k_0 \tau_0 -k_1\tau_1+i\varepsilon_0+i\varepsilon_1)}\,
e^{k_0\cdot k_1 |\tau_0-\tau_1| +i(\varepsilon_1\cdot k_0-\varepsilon_0\cdot k_1){\rm sgn}(\tau_0-\tau_1) +2\varepsilon_0\cdot\varepsilon_1\delta(\tau_0-\tau_1)}\Big|_{\rm m.l.}~,
\end{align}
where the $\delta(\tau_0-\tau_1)$ part yields the seagull diagram, whereas the time ordered parts ($\tau_0>\tau_1$ and $\tau_0<\tau_1$) yield the diagrams where photon and graviton are singly emitted by the scalar line. We thus get the following irreducible contribution to the Feynman amplitude
\begin{align}
&{\cal D}_{irred}^{(1,1)}(p,p';\varepsilon_1,k_1;\epsilon,k_0)=(p'^2+m^2)(p^2+m^2)\widetilde D^{(1,1)}(p,p'; \varepsilon_1,k_1;\epsilon,k_0)\nonumber\\&= e\kappa \Bigl[ (p-p')\cdot \epsilon \cdot \varepsilon_1 +\frac{\varepsilon_1\cdot p' p\cdot\epsilon\cdot p}{p\cdot k_0}
-\frac{\varepsilon_1\cdot p\, p'\cdot\epsilon\cdot p'}{p\cdot k_1} \Bigr]~.
\label{eq:one-one-irred}
\end{align}
Finally, let us consider the irreducible contribution to the two-photon one-graviton amplitude, which is obviously trickier than the previous ones, though the worldline approach allows to obtain a quite compact representation. We report here the final result (the interested reader will find details of the computation to the Appendix~\ref{sec:appendix})  which reads
\begin{align}
&{\cal D}_{irred}^{(2,1)}(p,p';\varepsilon_1,k_1,\varepsilon_2,k_2;\epsilon,k_0) = \kappa e^2\Bigg\{ 2(\varepsilon_1 \epsilon \varepsilon_2) -2 \frac{\varepsilon_1\cdot \varepsilon_2\, (p'\epsilon p')}{m^2+(p'+k_0)^2}-2 \frac{\varepsilon_1\cdot \varepsilon_2\, (p\epsilon p)}{m^2+(p+k_0)^2}\nonumber\\&
+2\frac{\varepsilon_1\cdot p\, (\varepsilon_2\epsilon (p'-p-k_1))}{m^2+(p+k_1)^2}+ 2\frac{\varepsilon_1\cdot p'\, (\varepsilon_2 \epsilon (p-p'-k_1))}{m^2+(p'+k_1)^2} \nonumber\\&
+2\frac{\varepsilon_2\cdot p\, (\varepsilon_1 \epsilon (p'-p-k_2))}{m^2+(p+k_2)^2}+ 2\frac{\varepsilon_2\cdot p'\, (\varepsilon_1 \epsilon (p-p'-k_2))}{m^2+(p'+k_2)^2} \nonumber\\&
+4\frac{(p'\epsilon p')\, \varepsilon_1\cdot (p+k_2)\, \varepsilon_2\cdot p}{((p+k_2)^2+m^2)((p'+k_0)^2+m^2)} +4\frac{(p'\epsilon p')\, \varepsilon_2\cdot (p+k_1)\, \varepsilon_1\cdot p}{((p+k_1)^2+m^2)((p'+k_0)^2+m^2)} \nonumber\\
&+ 4\frac{(p\epsilon p)\, \varepsilon_1\cdot (p'+k_2)\, \varepsilon_2\cdot p'}{((p+k_0)^2+m^2)((p'+k_2)^2+m^2)}+4\frac{(p\epsilon p)\, \varepsilon_2\cdot (p'+k_1)\, \varepsilon_1\cdot p'}{((p+k_0)^2+m^2)((p'+k_1)^2+m^2)} \nonumber\\
&+4\frac{((p+k_1)\epsilon (p'+k_2))\, \varepsilon_1\cdot p\, \varepsilon_2\cdot p'}{((p+k_1)^2+m^2)((p'+k_2)^2+m^2)}+4\frac{((p+k_2)\epsilon (p'+k_1))\, \varepsilon_2\cdot p\, \varepsilon_1\cdot p'}{((p+k_2)^2+m^2)((p'+k_1)^2+m^2)}\Bigg\}~.
\end{align}
In the next section we tackle the reducible part of the amplitude.

%%%%%%%%%%%%%%%%%%%%%%%%%%%%%%%%%%
\subsection{Reducible part of the amplitude}
\label{sec:red-part}
\begin{figure}[htbp]
\begin{center}
 \includegraphics[width=0.7\textwidth]{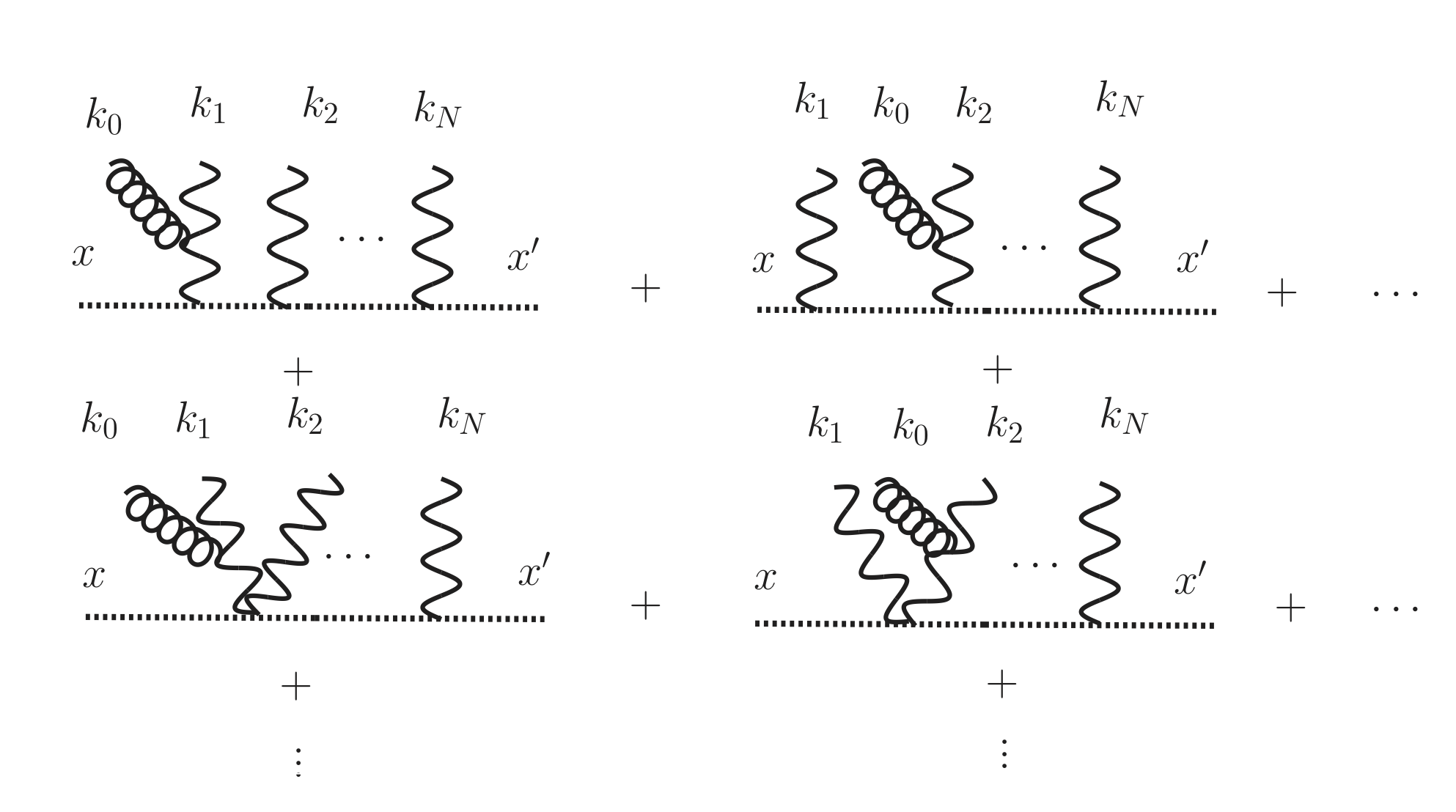}
\caption{Feynman diagram representation of the reducible contribution to $N$-photon one-graviton amplitude.}
\label{1grNph-red}
\end{center}
\end{figure}

The external graviton can couple directly to the scalar line, as reproduced by the formulas described in the previous section, but it can also couple to the photon lines---see Fig.~\ref{1grNph-red} for the  diagrammatic representation of these contributions. From a field theory view point this is encoded in the vertex
\begin{align}
{\cal V}[A,h] =\frac{\kappa}{2}\int d^4x h_{\mu\nu} T^{\mu\nu} = \frac{\kappa}{2}\int d^4x\, h_{\mu\nu} \Big( F^{\mu\alpha} F^\nu{}_\alpha -\frac14 \delta^{\mu\nu} F^{\alpha\beta}F_{\alpha\beta}\Big)\,,
\end{align} 
which, using the tracelessness of the on-shell graviton, leads to the following tree-level amplitude between two photons and one graviton
\begin{align}
\Gamma_{g\gamma\gamma}[\varepsilon,k,\varepsilon',k';\epsilon,k_0]=\kappa\Big[ (k\epsilon k) \varepsilon\cdot \varepsilon' + (\varepsilon \epsilon \varepsilon') k\cdot k_0- (\varepsilon\epsilon k) k\cdot \varepsilon'-(k \epsilon \varepsilon') \varepsilon \cdot k_0\Big]\,,
\label{eq:ggg}
\end{align} 
where $(a\epsilon b):=a_\mu \epsilon^{\mu\nu}b_\nu$, and we have used the  transversality conditions $k_{0\mu} \epsilon^{\mu\nu}=k_{\mu} \varepsilon^\mu =0$ and conservation law $k'=-(k+k_0)$. The latter can be used to construct the reducible part of the amplitude with the following recipe. Let us start from the one-photon two-scalar amplitude
\begin{align}
	{\cal D}^{(1)}(p,p';\varepsilon',k')=e \varepsilon'\cdot (p'-p)~,
\label{eq:one-red}
\end{align}
which can be easily read off from~\eqref{eq:MA}. It yields the reducible part of the one-photon one-graviton two-scalar amplitude by simply multiplying expressions~\eqref{eq:ggg} and~\eqref{eq:one-red}, and using the replacement rule
\begin{align}
	\varepsilon'^\alpha \varepsilon'^\beta\ \longrightarrow\ \frac{\delta^{\alpha\beta}}{k'^2}\,, 
\end{align}
which is the photon propagator in the Feynman gauge. By renaming photon polarization and momentum as $\varepsilon_1$ and $k_1$, we thus get
\begin{align}
	{\cal D}^{(1,1)}_{red}(p,p';\varepsilon_1,k_1;\epsilon,k_0)&= e\kappa (p'-p)_\mu \frac{\varepsilon_1^\mu (k_1 \epsilon k_1)
	+(\varepsilon_1 \epsilon)^\mu k_1\cdot k_0 -k_1^\mu (\varepsilon_1\epsilon k_1)  -(k_1 \epsilon)^\mu \varepsilon_1\cdot k_0}{2k_1\cdot k_0}~.
	\label{eq:one-one-red}
\end{align}
In other words we can obtain the latter as
\begin{align}
{\cal D}^{(1,1)}_{red}(p,p';\varepsilon_1,k_1;\epsilon,k_0)= {\cal D}^{(1)}(p,p';\upsilon_1,k_1+k_0)\,,
\end{align}
i.e., by starting from~\eqref{eq:one-red} and performing the replacement
\begin{align}
& \varepsilon^\mu_1\ \to \ \upsilon_1^\mu := \kappa  \frac{\varepsilon_1^\mu (k_1 \epsilon k_1)+(\varepsilon_1 \epsilon)^\mu k_1\cdot k_0 -k_1^\mu (\varepsilon_1\epsilon k_1)  -(k_1 \epsilon)^\mu \varepsilon_1\cdot k_0}{2k_1\cdot k_0} \label{eq:upsilon}\,, \\
& k_1^\mu \ \to \ k_1^\mu+k_0^\mu~,
\end{align}
note that~\eqref{eq:upsilon} is transversal upon the replacement $\varepsilon_1\ \to \ k_1$.
 The rule above can be obviously extended to the $N$-photon two-scalar amplitude constructed above in~\eqref{eq:MA}, which thus yields the following reducible contribution 
\begin{align}
{\cal D}^{(N,1)}_{red}(p,p';\varepsilon_1,k_1,\dots, \varepsilon_N,k_N;\epsilon,k_0) = \sum_{i=1}^N
{\cal D}^{(N)}(p,p';\varepsilon_1,k_1,\dots,\upsilon_i,k_i+k_0,\dots \varepsilon_N,k_N) ~.
\label{eq:red-part}
\end{align}
Thus, the full tree-level amplitude with $N$ photons, one graviton and two scalars reads
\begin{align}
{\cal D}^{(N,1)}(p,p';\varepsilon_1,k_1,\dots, \varepsilon_N,k_N;\epsilon,k_0)&={\cal D}^{(N,1)}_{irred}(p,p';\varepsilon_1,k_1,\dots, \varepsilon_N,k_N;\epsilon,k_0) \nonumber\\&\hspace{-1cm}+\sum_{l=1}^N
{\cal D}^{(N)}(p,p';\varepsilon_1,k_1,\dots,\upsilon_l,k_l+k_0,\dots \varepsilon_N,k_N)\,,
\label{eq:master}
\end{align}
where ${\cal D}^{(N,1)}_{irred}$ is given by eq.~\eqref{eq:tildeAg} `truncated' on the external scalar lines. For completeness, let us give the explicit expression for the reducible part of the amplitude with two photons. Let us start from the scalar Compton scattering amplitude, which can be easily obtained from ~\eqref{eq:MA} and reads
\begin{align}
{\cal D}^{(2)}(p,p';\varepsilon_1,k_1,\varepsilon_2,k_2)&= (-ie)^2\Big\{ 2\varepsilon_1\cdot \varepsilon_2 -\frac{\varepsilon_1\cdot (p'-p-k_2) \varepsilon_2\cdot (p'-p+k_1)}{(p'+k_1)^2+m^2}\nonumber\\&
-\frac{\varepsilon_1\cdot (p'-p+k_2) \varepsilon_2\cdot (p'-p-k_1)}{(p'+k_2)^2+m^2}\Bigg\}~.
\label{comptonqed}
\end{align}
By applying the replacement rule given above we get
\begin{align}
&{\cal D}^{(2,1)}_{red}(p,p';\varepsilon_1,k_1,\varepsilon_2,k_2;\epsilon,k_0) = {\cal D}^{(2)}(p,p';\upsilon_1,k_1+k_0,\varepsilon_2,k_2) +{\cal D}^{(2)}(p,p';\varepsilon_1,k_1,\upsilon_2,k_2+k_0)\nonumber\\
&= \kappa (-ie)^2 \Bigg\{\frac{2}{(k_1+k_0)^2}\big(\varepsilon_{1}^{\mu}\left(k_{1} \epsilon k_{1}\right)+\left(\varepsilon_{1} \epsilon\right)^{\mu} k_{1} \cdot k_{0}-k_{1}^{\mu}\left(\varepsilon_{1} \epsilon k_{1}\right)-\left(k_{1} \epsilon\right)^{\mu} \varepsilon_{1} \cdot k_{0} \big)\varepsilon_{2\mu} \nonumber\\
& -\frac{\varepsilon_2\cdot (p'-p+k_1+k_0)}{(p'+k_1+k_0)^2+m^2}\, \frac{\varepsilon_{1}^{\mu}\left(k_{1} \epsilon k_{1}\right)+\left(\varepsilon_{1} \epsilon\right)^{\mu} k_{1} \cdot k_{0}-k_{1}^{\mu}\left(\varepsilon_{1} \epsilon k_{1}\right)-\left(k_{1} \epsilon\right)^{\mu} \varepsilon_{1} \cdot k_{0}}{(k_1+k_0)^2} \, (p'-p-k_2)_\mu\nonumber\\
& -\frac{\varepsilon_2\cdot (p'-p-k_1-k_0)}{(p'+k_2)^2+m^2}\, \frac{\varepsilon_{1}^{\mu}\left(k_{1} \epsilon k_{1}\right)+\left(\varepsilon_{1} \epsilon\right)^{\mu} k_{1} \cdot k_{0}-k_{1}^{\mu}\left(\varepsilon_{1} \epsilon k_{1}\right)-\left(k_{1} \epsilon\right)^{\mu} \varepsilon_{1} \cdot k_{0}}{(k_1+k_0)^2} \, (p'-p+k_2)_\mu
\nonumber\\ & +(1\leftrightarrow 2)\Bigg\}~.
\label{eq:red-2-1}
\end{align}
Below, in Section~\ref{sec:WI-trans}, we test the master formula~\eqref{eq:master}  by checking the on-shell transversality conditions in the photon lines and graviton line. 
However, to conclude the present section, let us briefly review a factorization property that links graviton-photon amplitudes to photon amplitudes.
\subsection{On-shell factorization property for the graviton photoproduction  amplitude}
For a mixed scattering with one graviton and one photon, i.e. for the graviton photoproduction process, the full amplitude involving both the irreducible contributions~\eqref{eq:one-one-irred}  and the reducible contribution~\eqref{eq:one-one-red}, on-shell factorizes in terms of the corresponding QED Compton amplitude. It can be easily seen by adopting the decomposition
\begin{align}
	\epsilon^{\mu\nu}\rightarrow \epsilon^\mu\epsilon^\nu\,,
\end{align}
which yields,
%We can also look at the factorization property of the one-photon one-graviton scattering (linear gravitational Compton scattering) after adding both irreducible and reducible contributions. Putting together Eqs. (\ref{eq:one-one-irred}) and (\ref{eq:one-one-red}) and after some simple algebra (we decompose the graviton polarization as $\epsilon^{\mu\nu}\rightarrow \epsilon^\mu\epsilon^\nu$)
\begin{align}
\mathcal{M}^{(1,1)}(p,p';\varepsilon_1,k_1;\epsilon,k_0)
&=\frac{\kappa e}{k_0\cdot k_1}\Big[\epsilon\cdot p'\, k_0\cdot p-\epsilon\cdot p k_0\cdot p'\Big]\,\Big[\frac{\varepsilon_1\cdot p'\,\epsilon\cdot p}{p'\cdot k_1}+\frac{\varepsilon_1\cdot p\,\epsilon\cdot p'}{p'\cdot k_0}+\epsilon\cdot \varepsilon_1\Big]\nonumber\\
&=H{\cal M}^{(2)}(p,p';\epsilon,k_0,\varepsilon_1,k_1)\,,
\end{align}
where 
\begin{align}
H=-\frac{\kappa}{2e}\frac{\epsilon\cdot p'\, k_0\cdot p-\epsilon\cdot p k_0\cdot p'}{k_0\cdot k_1}\,,
\end{align}
and $\mathcal{M}^{(1,1)}(p,p';\varepsilon_1,k_1;\epsilon,k_0)$ and ${\cal M}^{(2)}(p,p';\epsilon,k_0,\varepsilon_1,k_1)$ are respectively the on-shell versions of the graviton photoproduction amplitude and of the scalar QED Compton scattering given in Eq. (\ref{comptonqed}). This factorization property was already studied in \cite{geohal-81,chshso-95,Holstein:2006ry,Bjerrum-Bohr:2014lea,Ahmadiniaz:2016vai}, and  seems to be universal for four-body amplitudes with massless gauge bosons. However, beyond the four-particle level, such factorization property is not expected to hold due to the lack of enough conservation laws~\cite{geohal-81}.

%%%%%%%%%%%%%%%%%%%%%%%%%%%%%%%%%%%%%%
\section{Ward identities and on-shell transversality}
\label{sec:WI-trans}

The dressed propagator described above in~\eqref{eq:dressed-prop}   is covariant upon $U(1)$ gauge transformations and invariant under diffeomorphisms. The former is described by
\begin{align}
	\Big\langle \phi(x')\bar\phi(x)\Big\rangle_{A,g}\ \to \ \Big\langle \tilde\phi(x')\tilde{\bar \phi}(x)\Big\rangle_{\tilde A, \tilde g}= e^{ie(\alpha(x)-\alpha(x'))} \Big\langle \phi(x')\bar\phi(x)\Big\rangle_{A,g}~.
	\label{eq:gauge-transf}
\end{align}
Using that $\delta A_\mu=\partial_\mu \alpha$, the infinitesimal part of~\eqref{eq:gauge-transf} becomes the electromagnetic Ward identity generator
\begin{align}
	\Big[\partial^y_\mu \frac{\delta}{\delta A_\mu(y)}+ie(\delta(y-x)-\delta(y-x'))\Big] \Big\langle \phi(x')\bar\phi(x)\Big\rangle_{A,g}=0~,
\label{eq:em-WI}
\end{align}
which holds off-shell. In momentum space, it yields an infinite set of  Ward identities
\begin{align}
\tilde D^{(N,1)}(p,p';-ik,k,\varepsilon_1,k_1,\dots; \epsilon,k_0) &= -ie\Big[
\tilde D^{(N-1,1)}(p+k,p';\varepsilon_1,k_1,\dots; \epsilon,k_0)\nonumber\\&
-\tilde D^{(N-1,1)}(p,p'+k;\varepsilon_1,k_1,\dots; \epsilon,k_0)\Big]\,,
\end{align}
which can be easily tested with the special cases singled out  in the section~\ref{sec:irred-part}.
On the other hand, on the scalar mass-shell the contact terms present in~\eqref{eq:em-WI} do not have the correct pole structure and drop out upon truncation, whereas the first term leads to the on-shell transversality condition
\begin{align}
	{\cal M}^{(N,1)}_{irred}(p,p';\varepsilon_1,k_1,\dots, -ik_l,k_l,\dots;\epsilon,k_0)=0~,
\end{align}
which holds for any photon line. As before ${\cal M}$ is the on-shell limit of ${\cal D}$. Moreover, the gauge invariance of scalar QED (in curved space) ensures that the full amplitude is transversal, i.e. the reducible part of the amplitude must result separately transversal. Indeed,  given that~\eqref{eq:upsilon} vanishes upon the replacement $\varepsilon_1\ \to\  k_1$, this is enough to prove the transversality of the reducible part of the amplitude~\eqref{sec:red-part}, as it can easily be checked for the expression~\eqref{eq:red-2-1}.  

Under infinitesimal diffeomorphisms, $x^\mu \to\ x^\mu -\xi^\mu(x)$, the dressed propagator transforms as
\begin{align}
\Big\langle \tilde\phi(x')\tilde{\bar \phi}(x)\Big\rangle_{\tilde A, \tilde g}=& \Big\langle \phi(x'){\bar \phi}(x)\Big\rangle_{ A, g} \nonumber\\&+\int d^4y\, \xi^\mu(y)\big( \delta^{(4)}(y-x)\partial_\mu+ \delta^{(4)}(y-x')\partial'_\mu\big)\Big\langle \phi(x'){\bar \phi}(x)\Big\rangle_{ A, g}\,.
\end{align}
However, using the worldline representation~\eqref{eq:dressed-prop}, one can as well get
\begin{align}
&\Big\langle \tilde\phi(x')\tilde{\bar \phi}(x)\Big\rangle_{\tilde A, \tilde g}= \Big\langle \phi(x'){\bar \phi}(x)\Big\rangle_{ A, g} \nonumber\\
&+\int d^4y\Big[ 2\nabla_\mu \xi_\nu(y) \frac{\delta}{\delta g_{\mu\nu}(y)} +\big(
\xi^\alpha\partial_\alpha A_\mu(y) +\partial_\mu \xi^\alpha A_\alpha(y)\big) \frac{\delta}{\delta A_{\mu}(y)}\Big] \Big\langle \phi(x'){\bar \phi}(x)\Big\rangle_{ A, g} ~,
\end{align}
which, after some straightforward algebra and using expression~\eqref{eq:em-WI}, can be reduced to
\begin{align}
	&\Biggl[-\nabla^y_\mu \frac{2g_{\nu\alpha}}{\sqrt{g}}\frac{\delta}{\delta g_{\mu\nu}(y)}\nonumber\\&
	+\frac{1}{\sqrt{g}}\Big(F_{\alpha\mu}\frac{\delta}{\delta A_\mu(y)} -\delta^{(4)}(y-x) \bar D_\alpha -\delta^{(4)}(y-x') D_\alpha'\Big)\Biggr] \Big\langle \phi(x')\bar\phi(x)\Big\rangle_{A,g}=0~,
\label{eq:diff-WI}
\end{align}
which is the diffeomorphism Ward identity generator. Once again there are  contact terms which drop out on the scalar particle mass-shell. The two left-over terms both contribute on-shell and thus the irreducible part of the $N$-photon one-graviton amplitude is not, by itself, transversal on the graviton line; rather it fulfills, even on-shell, an inhomogeneous Ward identity. 
Introducing the field strength tensor 
$f_i^{\mu\nu} := k_i^{\mu}\varepsilon^{\nu}_i - \varepsilon^{\mu}_ik_i^{\nu}$
for each photon leg, and an ``effective'' photon polarization vector
\begin{eqnarray}
\tilde\varepsilon_i := \kappa f_i\cdot \xi \,,
\label{defepsilontilde}
\end{eqnarray}
this identity can be written concisely as follows (the same identity holds for the 
closed-loop case \cite{Bastianelli:2012bz})
\begin{eqnarray}
\tilde D^{(N,1)}(p,p';\varepsilon_1,k_1,\dots; k_0\xi,k_0) = 
\sum_{i=1}^N \tilde D^{(N,0)}(p,p';\varepsilon_1,k_1,\dots,\tilde\varepsilon_i,k_i + k_0, \ldots, \varepsilon_N,k_N)\,.
\end{eqnarray}
Here we have written the transformation of the (transverse traceless) polarization tensor as
\begin{align}
\epsilon_{\mu\nu} \ \to\ \epsilon_{\mu\nu} +k_{0\mu} \xi_\nu + k_{0\nu} \xi_\mu~,\quad k_0\cdot \xi =k_0^2=0~, 
\label{eq:trans-g}
\end{align}
and used $k_0\xi$ just a shortcut notation for the symmetrized product of the two vectors.
However, the full amplitude is expected to be transversal on-shell, i.e.,
 \begin{align}
{\cal M}^{(N,1)}(p,p';\varepsilon_1,k_1,\dots, \varepsilon_N,k_N;k_0\xi,k_0)&= 0~ \, .
\end{align}
Using the ``tree replacement'' rule \eqref{eq:upsilon}, it can be seen quite easily how this comes about:
applying the transformation \eqref{eq:trans-g} to $\upsilon_i^{\mu}$, the result can be written as
\begin{equation}
\upsilon_i^{\mu} \to  -\tilde \varepsilon_i^{\mu} +
\kappa\frac{k_0\cdot f_i \cdot \xi}{2k_i\cdot k_0} (k_0+k_i)^{\mu} \, .
\end{equation}
The second term in brackets will drop out when inserted into the photon amplitude
because of the transversality in the photon polarizations. The first one will cancel the
contribution of the $i$th term on the right-hand side of \eqref{eq:red-part} to the
Ward identity. In the Appendix~\ref{sec:examples} we single out a few detailed examples.

%%%%%%%%%%%%%%%%
\section{Conclusions and Outlook}
\label{sec:concl}
We described a novel worldline approach to the computation of the tree level scattering amplitudes associated to the scalar line coupled to electromagnetism and gravity with all external legs  off-shell. In particular, we provided a convenient parametrization for the graviton polarization and a replacement rule, which allowed us to easily compute amplitudes with an arbitrary number of photons and one graviton. The on-shell transversality of the amplitudes was explicitly checked.  

A priori, our technique can be as well implemented to compute amplitudes with an arbitrary number of gravitons. However, in that case more care is needed in the treatment of chains of contractions between the Lee-Yang ghost fields that represent the non trivial measure~\cite{Bastianelli:1991be, Bastianelli:1992ct}.

On the other hand amplitudes with gravitons have always been the subject of extensive studies. In particular, theorems which involve gravitons with low momentum have long been analyzed~\cite{Weinberg:1964ew} and, in the recent past, various soft-graviton theorems---see e.g. Ref.~\cite{Cachazo:2014fwa}---have been studied, due to their connections to the infrared structure of gauge theory and gravity~\cite{Strominger:2017zoo}. The present manuscript wishes to provide a novel approach towards the computation of amplitudes with gravitons, which may shed new light on the structure of such quantities. In fact, our approach does not, a priori, require gravitons to have low-momentum. However, it would be helpful to reconstruct soft graviton theorems from the worldline view point, by suitably implementing from the beginning the low-momentum condition into the graviton vertex operators~\eqref{eq:grav-vert-op}. Yet, the parametrization described in Section~\ref{sec:irred-part}, which allows to simplify the computation of the worldline correlators, keeps holding for each graviton vertex operator.

%%%%%%%%%%%%%%%%%%%%%%
\subsection*{Ackowledgments}
The Authors would like to thank Fiorenzo Bastianelli, James Edwards and Diego Trancanelli for helpful discussions.

%%%%%%%%%%%%%%%%%%%%%%%%%
\begin{appendices}

\section{Two-photon one-graviton scalar propagator}
\label{sec:appendix}
\begin{figure}[htbp]
\begin{center}
 \includegraphics[width=0.7\textwidth]{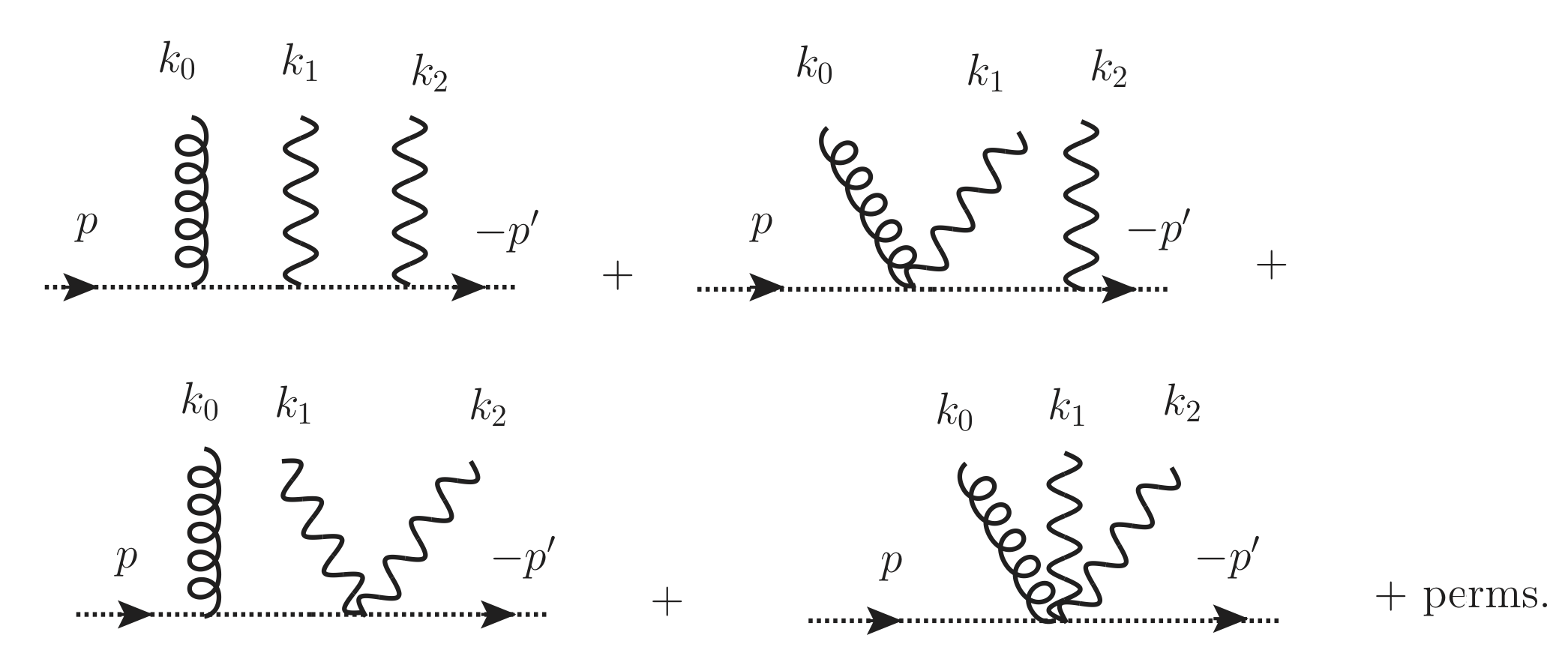}
\caption{Irreducible contributions to the amplitude with two-photon one-graviton, which are shown here in momentum space. 'Perms' refers to permutations between the photon lines and among the emission points. The last type of diagram, where photons and graviton are all emitted at the same point, is obviously unique.}
\label{2ph1gr}
\end{center}
\end{figure}

We use the master formula~\eqref{eq:tildeAg} to compute the  two-photon one-graviton scalar propagator, and the related (irreducible) part of the two-photon one-graviton two-scalar amplitude, whose Feynman diagrams are depicted in~Fig~\ref{2ph1gr}. 

\noindent It reduces to,
\begin{align}
&\widetilde D^{(2,1)}(p,p'; \varepsilon_1,k_1, \varepsilon_2, k_2;\epsilon,k_0)=(-ie)^2\left(-\frac{\kappa}{4}\right)\int_0^\infty dT ~e^{-T(m^2+p'^2)}\int_0^T d\tau_0 \int_0^T d\tau_1\int_0^T d\tau_2\nonumber\\
&\times e^{(p'-p)\cdot (-k_0 \tau_0 -k_1\tau_1-k_2\tau_2+i\varepsilon_0+i\varepsilon_1+i\varepsilon_2)}\,
e^{k_0\cdot k_1 |\tau_0-\tau_1| +k_0\cdot k_2 |\tau_0-\tau_2| +k_1\cdot k_2 |\tau_1-\tau_2|} \nonumber\\&\times
e^{i(\varepsilon_1\cdot k_0-\varepsilon_0\cdot k_1){\rm sgn}(\tau_0-\tau_1)+i(\varepsilon_2\cdot k_0-\varepsilon_0\cdot k_2){\rm sgn}(\tau_0-\tau_2)+i(\varepsilon_2\cdot k_1-\varepsilon_1\cdot k_2){\rm sgn}(\tau_1-\tau_2) }\nonumber\\&
\times e^{2\big[\varepsilon_0\cdot\varepsilon_1\delta(\tau_0-\tau_1)+\varepsilon_0\cdot\varepsilon_2\delta(\tau_0-\tau_2)
+\varepsilon_1\cdot\varepsilon_2\delta(\tau_1-\tau_2)\big]}\Big|_{\rm m.l.}~.
\end{align}
Firstly, let us consider contributions involving delta functions, which are linked to seagull diagrams. We find it convenient to  `grade' the different contributions in terms of how many delta functions occur. There is only one double-delta term (see the last diagram in Fig. \ref{2ph1gr}), namely
\begin{align}
e^2\kappa
\int_0^\infty dT\, e^{-T(m^2+p'^2)} \int_0^T d\tau_0~ e^{\tau_0 (p'^2-p^2)}\, \varepsilon_0\cdot \varepsilon_1 \varepsilon_0\cdot\varepsilon_2|_{\rm m.l.} ~,
\end{align}
which, using~\eqref{eq:e} and~\eqref{eq:lr}, reduces to
\begin{align}
 \frac{1}{(p^2+m^2)(p'^2+m^2)} \, e^2\kappa\, 2(\varepsilon_1 \epsilon \varepsilon_2)~,
\end{align} 
whose numerator is the Feynman amplitude of the diagram where two photons and one graviton are emitted at the same point of the scalar line. Note that, also for an arbitrary number $N$ of photons---and a single graviton---this is the largest number of particles that can be emitted at the same point of the scalar line.

There are three terms with a single delta function (see the second and third diagrams and their permutations in Fig. \ref{2ph1gr}), that correspond to the six Feynman diagrams where there is the emission of a pair of particles (either two photons or one photon and the graviton) from the same point of the scalar line, and the remaining particle emitted from another point on the line. Let us, for example consider the term that involves $\delta(\tau_1-\tau_2)$ (the third diagram in Fig. \ref{2ph1gr}), which yield the diagrams where two photons are emitted at the same point. The integrand reads
\begin{align}
	&(-ie)^2\left(-\frac{\kappa}{4}\right)\varepsilon_1 \cdot\varepsilon_2 \, \left[  i\varepsilon_0 \cdot(p'-p-(k_1+k_2){\rm sgn}(\tau_0-\tau_1))\right]^2\nonumber\\
	&\times e^{(p-p')\cdot(k_0\tau_0+(k_1+k_2)\tau_1)+k_0\cdot(k_1+k_2)|\tau_0-\tau_1|}~,
\end{align}
which provides two diagrams, according to whether $\tau_1<\tau_0$
 or $\tau_0<\tau_1$. After some straightforward algebra that corresponds to the Schwinger integral parametrization of the diagrams, we obtain
 \begin{align}
 	\frac1{(m^2+p^2)(m^2+p'^2)}(-2e^2 \kappa\varepsilon_1\cdot \varepsilon_2)\Big[\frac{(p'\epsilon p')}{m^2+(p'+k_0)^2}+ \frac{(p\epsilon p)}{m^2+(p+k_0)^2}\Big]~.
 \end{align}
Similarly, the other terms with single delta functions $\delta(\tau_0-\tau_1)$ and $\delta(\tau_0-\tau_2)$ give
\begin{align}
	\frac1{(m^2+p^2)(m^2+p'^2)}\, 2e^2 \kappa\Big[&\frac{\varepsilon_1\cdot p\, (\varepsilon_2 \epsilon (p'-p-k_1))}{m^2+(p+k_1)^2}+ \frac{\varepsilon_1\cdot p'\, (\varepsilon_2 \epsilon (p-p'-k_1))}{m^2+(p'+k_1)^2}\nonumber\\
	& +\frac{\varepsilon_2\cdot p\, (\varepsilon_1 \epsilon (p'-p-k_2))}{m^2+(p+k_2)^2}+ \frac{\varepsilon_2\cdot p'\, (\varepsilon_1 \epsilon (p-p'-k_2))}{m^2+(p'+k_2)^2} \Big]~.
\end{align}

The term without delta functions corresponds to the leftover six Feynman diagrams where the two photons and the graviton and emitted singly by the scalar line (the first diagram in Fig. \ref{2ph1gr} and its permutations), six being the number of permutations of the three particles, which in the present worldline representation correspond to the different orderings of the three times $\tau_i$. The integrand in this case reads
\begin{align}
	&(-ie)^2\left(-\frac{\kappa}{4}\right)\int_0^\infty dT ~e^{-T(m^2+p'^2)}\int_0^T d\tau_0 \int_0^T d\tau_1\int_0^T d\tau_2\,\nonumber\\
&\times e^{(p'-p)\cdot (-k_0 \tau_0 -k_1\tau_1-k_2\tau_2+i\varepsilon_0+i\varepsilon_1+i\varepsilon_2)}\,
e^{k_0\cdot k_1 |\tau_0-\tau_1| +k_0\cdot k_2 |\tau_0-\tau_2| +k_1\cdot k_2 |\tau_1-\tau_2|} \nonumber\\&\times  \varepsilon_1\cdot (p'-p+k_0{\rm sgn}(\tau_0-\tau_1) -k_2{\rm sgn}(\tau_1-\tau_2))\,
\nonumber\\&\times \varepsilon_2\cdot (p'-p+k_0{\rm sgn}(\tau_0-\tau_2) +k_1{\rm sgn}(\tau_1-\tau_2))\nonumber\\
&\times 
\frac12 \big[\varepsilon_0\cdot (p'-p-k_1{\rm sgn}(\tau_0-\tau_1) -k_2{\rm sgn}(\tau_0-\tau_2))\big]^2\,,
\end{align}
and yields
\begin{align}
\frac1{(m^2+p^2)(m^2+p'^2)}\, 4 e^2 \kappa\Big[ &\frac{(p'\epsilon p')\, \varepsilon_1\cdot (p+k_2)\, \varepsilon_2\cdot p}{((p+k_2)^2+m^2)((p'+k_0)^2+m^2)} +(1\leftrightarrow 2)\nonumber\\
+& \frac{(p\epsilon p)\, \varepsilon_1\cdot (p'+k_2)\, \varepsilon_2\cdot p'}{((p+k_0)^2+m^2)((p'+k_2)^2+m^2)} +(1\leftrightarrow 2)\nonumber\\
+&\frac{((p+k_1)\epsilon (p'+k_2))\, \varepsilon_1\cdot p\, \varepsilon_2\cdot p'}{((p+k_1)^2+m^2)((p'+k_2)^2+m^2)} +(1\leftrightarrow 2)\Big]~.
\end{align}
Thus,
\begin{align}
&{\cal D}_{irred}^{(2,1)}(p,p';\varepsilon_1,k_1,\varepsilon_2,k_2;\epsilon,k_0) = \kappa e^2\Bigg\{ 2(\varepsilon_1 \epsilon \varepsilon_2) -2 \frac{\varepsilon_1\cdot \varepsilon_2\, (p'\epsilon p')}{m^2+(p'+k_0)^2}-2 \frac{\varepsilon_1\cdot \varepsilon_2\, (p\epsilon p)}{m^2+(p+k_0)^2}\nonumber\\&
+2\frac{\varepsilon_1\cdot p\, (\varepsilon_2\epsilon (p'-p-k_1))}{m^2+(p+k_1)^2}+ 2\frac{\varepsilon_1\cdot p'\, (\varepsilon_2 \epsilon (p-p'-k_1))}{m^2+(p'+k_1)^2} \nonumber\\&
+2\frac{\varepsilon_2\cdot p\, (\varepsilon_1 \epsilon (p'-p-k_2))}{m^2+(p+k_2)^2}+ 2\frac{\varepsilon_2\cdot p'\, (\varepsilon_1 \epsilon (p-p'-k_2))}{m^2+(p'+k_2)^2} \nonumber\\&
+4\frac{(p'\epsilon p')\, \varepsilon_1\cdot (p+k_2)\, \varepsilon_2\cdot p}{((p+k_2)^2+m^2)((p'+k_0)^2+m^2)} +4\frac{(p'\epsilon p')\, \varepsilon_2\cdot (p+k_1)\, \varepsilon_1\cdot p}{((p+k_1)^2+m^2)((p'+k_0)^2+m^2)} \nonumber\\
&+ 4\frac{(p\epsilon p)\, \varepsilon_1\cdot (p'+k_2)\, \varepsilon_2\cdot p'}{((p+k_0)^2+m^2)((p'+k_2)^2+m^2)}+4\frac{(p\epsilon p)\, \varepsilon_2\cdot (p'+k_1)\, \varepsilon_1\cdot p'}{((p+k_0)^2+m^2)((p'+k_1)^2+m^2)} \nonumber\\
&+4\frac{((p+k_1)\epsilon (p'+k_2))\, \varepsilon_1\cdot p\, \varepsilon_2\cdot p'}{((p+k_1)^2+m^2)((p'+k_2)^2+m^2)}+4\frac{((p+k_2)\epsilon (p'+k_1))\, \varepsilon_2\cdot p\, \varepsilon_1\cdot p'}{((p+k_2)^2+m^2)((p'+k_1)^2+m^2)}\Bigg\}\,,
\end{align}
is the irreducible part of the two-scalar two-photon one-graviton amplitude.
\section{Transversality of the amplitudes with one graviton and $N\leq 2$ photons}\label{sec:examples}

Let us here check how the transversality of the graviton line explicitly works for $N\leq 2$. For the $N=0$ amplitude of eq.~\eqref{eq:g} we have
\begin{align}
{\cal M}^{(0,1)}(p,p';k_0\xi,k_0) =\frac{\kappa}{2} (p'-p)\cdot k_0\, (p'-p)\cdot \xi \,,
\end{align}
which vanishes on-sell because $k_0=-(p+p')$. For $N=1$, using on-shellness, the momentum conservation and the transversality conditions $k_{0\mu} \epsilon^{\mu\nu}=k_{\mu} \varepsilon^\mu =0$, we have
\begin{align}
{\cal M}^{(1,1)}_{red}(p,p';\varepsilon_1,k_1;k_0\xi,k_0) =-{\cal M}^{(1,1)}_{irred}(p,p';\varepsilon_1,k_1;k_0\xi,k_0) =e\kappa (p'-p)_\mu \Big( \varepsilon_1^\mu k_1\cdot \xi +k_0^\mu \varepsilon_1\cdot \xi\Big)\,,
\end{align} 
so that
\begin{align}
{\cal M}^{(1,1)}(p,p';\varepsilon_1,k_1;k_0\xi,k_0) =0\,,
\end{align}
as expected. 

The computation for the $N=2$ case is of course more complicated. However, let us sketch some details.  An useful way to proceed is to identify different \emph{kind} of terms in both the reducible and irreducible parts of the amplitude, that must sum up to zero separately. 

Let us first consider the part of the amplitude proportional to the product $\varepsilon_1\cdot\varepsilon_2$. After performing the substitution described in Eq.\eqref{eq:trans-g}, and denoting the corresponding reducible and irreducible contributions as $\mathcal{M}_{red}^{\varepsilon_1\varepsilon_2}$ and $\mathcal{M}_{irred}^{\varepsilon_1\varepsilon_2}$, we obtain
\begin{align}
\mathcal{M}_{irred}^{\varepsilon_1\varepsilon_2}=&-\frac{2\varepsilon_1\cdot\varepsilon_2}{p\cdot k_0}\left(p\cdot k_0 p\cdot\xi\right)-\frac{2\varepsilon_1\cdot\varepsilon_2}{p'\cdot k_0}\left(p'\cdot k_0 p'\cdot\xi\right)=-2\varepsilon_1\cdot\varepsilon_2\xi \cdot(p+p')\,,\\
\mathcal{M}_{red}^{\varepsilon_1\varepsilon_2}=&-\frac{2\varepsilon_1\cdot\varepsilon_2}{k_1\cdot k_0}\left(k_1\cdot k_0 k_1\cdot\xi\right)-\frac{2\varepsilon_1\cdot\varepsilon_2}{k_2\cdot k_0}\left(k_2\cdot k_0 k_2\cdot\xi\right)=\nonumber\\
=&-2\varepsilon_1\cdot\varepsilon_2\xi \cdot(k_1+k_2)=2\varepsilon_1\cdot\varepsilon_2\xi \cdot(p+p')=-\mathcal{M}_{irred}^{\varepsilon_1\varepsilon_2},
\end{align}
where in the last line we have used the conservation of total energy-momentum together with the transversality condition given in Eq.~\eqref{eq:trans-g}. Thus, we get
\begin{equation}
\mathcal{M}_{irred}^{\varepsilon_1\varepsilon_2}+\mathcal{M}_{red}^{\varepsilon_1\varepsilon_2}=0,
\end{equation}
as expected.\\
Similarly we could consider the part in the total amplitude proportional to $\varepsilon_1\cdot\xi$, and we indicate with $\mathcal{M}_{red}^{\varepsilon_1\xi}$ and $\mathcal{M}_{irred}^{\varepsilon_1\xi}$ respectively the reducible and irreducible contributions. After some manipulations, we obtain
\begin{align}
\mathcal{M}_{irred}^{\varepsilon_1\xi}=&\frac{\varepsilon_2\cdot p'}{k_2\cdot p'}p\cdot k_0 \varepsilon_1\cdot\xi +2\varepsilon_1\cdot\xi\varepsilon_2\cdot k_0+\frac{\varepsilon_2\cdot p'}{k_2\cdot p'}(p+k_1)\cdot k_0 \varepsilon_1\cdot\xi\nonumber\\
&+\frac{\varepsilon_2\cdot p}{k_2\cdot p}p'\cdot k_0 \varepsilon_1\cdot\xi +\frac{\varepsilon_2\cdot p}{k_2\cdot p}(p'+k_1)\cdot k_0 \varepsilon_1\cdot\xi\nonumber\\
=&\frac{\varepsilon_2\cdot p'}{k_2\cdot p'}p\cdot k_0 \varepsilon_1\cdot\xi +2\varepsilon_1\cdot\xi\varepsilon_2\cdot k_0-\frac{\varepsilon_2\cdot p'}{k_2\cdot p'}p\cdot k_1 \varepsilon_1\cdot\xi+\varepsilon_1\cdot\xi\varepsilon_2\cdot p'\nonumber\\
&+\frac{\varepsilon_2\cdot p}{k_2\cdot p}p'\cdot k_0 \varepsilon_1\cdot\xi-\frac{\varepsilon_2\cdot p}{k_2\cdot p}p'\cdot k_1 \varepsilon_1\cdot\xi+\varepsilon_1\cdot\xi\varepsilon_2\cdot p\nonumber\\
=&\frac{\varepsilon_2\cdot p'}{k_2\cdot p'}\varepsilon_1\cdot\xi p\cdot(k_0-k_1)+\frac{\varepsilon_2\cdot p}{k_2\cdot p} \varepsilon_1\cdot\xi p'\cdot(k_0-k_1)
+\varepsilon_1\cdot\xi\varepsilon_2\cdot (k_0-k_1).
\end{align}
Notice that in the last equality we have exploited the conservation of total energy-momentum, while in the second equality we have used the relations
\begin{align}\label{RAN}
&k_0\cdot(p+k_1)=-p\cdot k_1+p'\cdot k_2,\nonumber\\
&k_0\cdot(p'+k_1)=-p'\cdot k_1+p\cdot k_2.
\end{align}
The contribution coming from the reducible part of the amplitude is obtained as
\begin{align}
\mathcal{M}_{red}^{\varepsilon_1\xi}=&\frac{\varepsilon_2\cdot p'}{p'\cdot k_2k_0\cdot k_1}\varepsilon_1\cdot\xi (p\cdot k_1k_0\cdot k_1-p\cdot k_0k_0\cdot k_1)  +2\Bigg(\frac{\varepsilon_1\cdot \xi}{2k_0\cdot k_1}(k_0\cdot k_1\varepsilon_2\cdot k_1\nonumber\\
&-k_0\cdot k_1\varepsilon_2\cdot k_0)\Bigg)  +\frac{\varepsilon_2\cdot p}{p\cdot k_2k_0\cdot k_1}\varepsilon_1\cdot\xi (p'\cdot k_1k_0\cdot k_1-p'\cdot k_0k_0\cdot k_1)\nonumber\\
&=-\frac{\varepsilon_2\cdot p'}{k_2\cdot p'}\varepsilon_1\cdot\xi p\cdot(k_0-k_1)-\frac{\varepsilon_2\cdot p}{k_2\cdot p} \varepsilon_1\cdot\xi p'\cdot(k_0-k_1)
-\varepsilon_1\cdot\xi\varepsilon_2\cdot (k_0-k_1),
\end{align}
and the sum of the reducible and irreducible contribution vanishes, that is
\begin{equation}
\mathcal{M}_{irred}^{\varepsilon_1\xi}+\mathcal{M}_{red}^{\varepsilon_1\xi}=0.
\end{equation}
By Bose symmetry the contributions proportional to $\varepsilon_2\cdot\xi$ can be obtained from the latter with the replacements $\varepsilon_1\leftrightarrow\varepsilon_2$ and $k_1\leftrightarrow k_2$.
Now we are ready to write down all the remaining terms that enter in the transversality expression for the total amplitude. We find it convenient to organize them in terms of their different denominators, which are scalar product of momenta. We thus use the notation $\mathcal{M}_{rem}^{pk}$ to indicate those terms that have the common denominator $p\cdot k$ and similarly with others. We have,
\begin{align}
\mathcal{M}_{rem}^{p'k_2}=&-\frac{\varepsilon_2\cdot p'}{p'\cdot k_2}2p\cdot\xi\varepsilon_1\cdot(p+k_0)+\frac{\varepsilon_2\cdot p'}{p'\cdot k_2}\varepsilon_1\cdot k_0 p\cdot\xi+\frac{\varepsilon_2\cdot p'}{p'\cdot k_2}\varepsilon_1\cdot k_0\xi\cdot(p+k_1)\nonumber\\
&+\frac{\varepsilon_2\cdot p'}{p'\cdot k_2}2\varepsilon_1\cdot p\xi\cdot(p+k_1)-\frac{\varepsilon_2\cdot p'}{p'\cdot k_2}2p\cdot\varepsilon_1\xi\cdot k_1-\frac{\varepsilon_2\cdot p'}{p'\cdot k_2}\varepsilon_1\cdot k_0\xi\cdot k_1=0\,,
\end{align}
\begin{align}
\mathcal{M}_{rem}^{pk_1}=&-\frac{\varepsilon_1\cdot p}{p\cdot k_1}\varepsilon_2\cdot k_0\xi\cdot (p+k_1)-\frac{\varepsilon_1\cdot p}{p\cdot k_1}2p'\cdot \xi\varepsilon_2\cdot(p'+k_0)+\frac{\varepsilon_1\cdot p}{p\cdot k_1}\varepsilon_2\cdot k_0\xi\cdot p'\nonumber\\
&-\frac{\varepsilon_1\cdot p}{p\cdot k_1}2\varepsilon_2\cdot p'\xi\cdot(p+k_1)+\frac{\varepsilon_1\cdot p}{p\cdot k_1}2(p+k_1)\cdot\varepsilon_2\xi\cdot k_2+\frac{\varepsilon_1\cdot p}{p\cdot k_1}\varepsilon_2\cdot k_0\xi\cdot k_2\nonumber\\
=&\,\frac{\varepsilon_1\cdot p}{p\cdot k_1}\varepsilon_2\cdot k_0\xi\cdot (p'+k_2)-\frac{\varepsilon_1\cdot p}{p\cdot k_1}2p'\cdot \xi\varepsilon_2\cdot(p'+k_0)+\frac{\varepsilon_1\cdot p}{p\cdot k_1}\varepsilon_2\cdot k_0\xi\cdot p'\nonumber\\
&+\frac{\varepsilon_1\cdot p}{p\cdot k_1}2\varepsilon_2\cdot p'\xi\cdot(p'+k_2)-\frac{\varepsilon_1\cdot p}{p\cdot k_1}2(p'+k_0)\cdot\varepsilon_2\xi\cdot k_2+\frac{\varepsilon_1\cdot p}{p\cdot k_1}\varepsilon_2\cdot k_0\xi\cdot k_2=0\,,
\end{align}
\begin{align}
\mathcal{M}_{rem}^{p'k_1}=&-\frac{\varepsilon_1\cdot p'}{p'\cdot k_1}2p\cdot\xi\varepsilon_2\cdot(p+k_0)+\frac{\varepsilon_1\cdot p'}{p'\cdot k_1}p\cdot\xi\varepsilon_2\cdot k_0-\frac{\varepsilon_1\cdot p'}{p'\cdot k_1}\varepsilon_2\cdot k_0\xi\cdot(p'+k_1)\nonumber\\
&-\frac{\varepsilon_1\cdot p'}{p'\cdot k_1}2\varepsilon_2\cdot p\xi\cdot(p'+k_1)+\frac{\varepsilon_1\cdot p'}{p'\cdot k_1}2(p'+k_1)\cdot\varepsilon_2\xi\cdot k_2+\frac{\varepsilon_1\cdot p'}{p'\cdot k_1}\varepsilon_2\cdot k_0\xi\cdot k_2\nonumber\\
=&-\frac{\varepsilon_1\cdot p'}{p'\cdot k_1}2p\cdot\xi\varepsilon_2\cdot(p+k_0)+\frac{\varepsilon_1\cdot p'}{p'\cdot k_1}p\cdot\xi\varepsilon_2\cdot k_0+\frac{\varepsilon_1\cdot p'}{p'\cdot k_1}\varepsilon_2\cdot k_0\xi\cdot(p+k_2)\nonumber\\
&+\frac{\varepsilon_1\cdot p'}{p'\cdot k_1}2\varepsilon_2\cdot p\xi\cdot(p+k_2)-\frac{\varepsilon_1\cdot p'}{p'\cdot k_1}2(p+k_0)\cdot\varepsilon_2\xi\cdot k_2+\frac{\varepsilon_1\cdot p'}{p'\cdot k_1}\varepsilon_2\cdot k_0\xi\cdot k_2=0\,,
\end{align}
\begin{align}
\mathcal{M}_{rem}^{pk_2}=&-\frac{\varepsilon_2\cdot p}{p\cdot k_2}2p'\cdot \xi\varepsilon_1\cdot(p'+k_0)+\frac{\varepsilon_2\cdot p}{p\cdot k_2}\varepsilon_1\cdot k_0\xi\cdot p'+\frac{\varepsilon_2\cdot p}{p\cdot k_2}\varepsilon_1\cdot k_0\xi\cdot(p'+k_1)\nonumber\\
&+\frac{\varepsilon_2\cdot p}{p\cdot k_2}2\varepsilon_1\cdot p'\xi\cdot(p'+k_1)-\frac{\varepsilon_2\cdot p}{p\cdot k_2}2p'\cdot\varepsilon_1\xi\cdot k_1-\frac{\varepsilon_2\cdot p}{p\cdot k_2}\varepsilon_1\cdot k_0\xi\cdot k_1=0\,,
\end{align}
\begin{align}
\mathcal{M}_{rem}^{k_0k_1}=&\frac{\varepsilon_1\cdot k_0\xi\cdot k_1}{k_0\cdot k_1}\varepsilon_2\cdot(k_0+k_1)+\frac{\varepsilon_1\cdot k_0}{k_0\cdot k_1}\varepsilon_2\cdot p'\xi\cdot k_1+\frac{\varepsilon_1\cdot k_0}{k_0\cdot k_1}\varepsilon_2\cdot p\xi\cdot k_1\nonumber\\
=&\,\frac{\varepsilon_1\cdot k_0}{k_0\cdot k_1}\xi\cdot k_1\varepsilon_2\cdot (p+p'+k_0+k_1)\propto \varepsilon_2\cdot k_2=0\,,
\end{align}
\begin{align}
\mathcal{M}_{rem}^{k_0k_2}=&\frac{\varepsilon_2\cdot k_0\xi\cdot k_2}{k_0\cdot k_2}\varepsilon_1\cdot(k_0+k_2)+\frac{\varepsilon_2\cdot k_0}{k_0\cdot k_2}\varepsilon_1\cdot p\xi\cdot k_2+\frac{\varepsilon_2\cdot k_0}{k_0\cdot k_2}\varepsilon_1\cdot p'\xi\cdot k_2\nonumber\\
=&\,\frac{\varepsilon_2\cdot k_0}{k_0\cdot k_2}\xi\cdot k_2\varepsilon_1\cdot (p+p'+k_0+k_2)\propto \varepsilon_1\cdot k_1=0~.
\end{align}
Thus, all the different contributions sum up to zero and the \emph{transversality} of the total amplitude is proven, i.e.,
\begin{equation}
\mathcal{M}^{(2,1)}\left(p,p';\varepsilon_1,k_1;\varepsilon_2,k_2;k_0\xi,k_0\right)=0~.
\end{equation}
%So, once the substitution in Eq.(\ref{FTA}) is performed, is useful first of all to rewrite the contribution from the \emph{reducible} part of the amplitude, that we indicate as $\mathcal{M}_{red}^{(2,1)}$. This is given by the sum of four different terms, that is
%\begin{equation}
%\mathcal{M}_{red}^{(2,1)}=\mathcal{M}_{red,1}^{12}+\mathcal{M}_{red,2}^{12}+\mathcal{M}_{red,1}^{21}+\mathcal{M}_{red,2}^{21}.
%\end{equation}

 What we described above is similar to what happens in flat space scalar QCD, for which a worldline approach to the computation of the $N$-gluon scalar propagator was studied in~\cite{Ahmadiniaz:2015xoa}: it yields the irreducible part of the $N$-gluon two-scalar amplitude. However, the non-Abelian nature of the theory implies that in order to compute the full amplitude---which is guaranteed to be transversal on the gluon lines---the latter must be completed with reducible parts~\cite{FMB-thesis}. 

\end{appendices}
%%%%%%%%%%%%%%%%%%%%%

\end{document}